\documentclass[twocolumn,showpacs,amsmath,amssymb]{revtex4}

\usepackage{graphicx}
\usepackage{dcolumn}
\usepackage{bm}
\newcommand{\etal}{\emph{et al.}}
\newcommand{\be}{\begin{equation}}
\newcommand{\ee}{\end{equation}}
\newcommand{\bfig}{\begin{figure}}
\newcommand{\efig}{\end{figure}}

\begin{document}      
\title{Ultrahigh mobility and giant magnetoresistance in the Dirac semimetal Cd$_3$As$_2$.} 
 
\author{Tian Liang$^{1}$}
\author{Quinn Gibson$^{2}$}
\author{Mazhar N. Ali$^2$}
\author{Minhao Liu$^1$}
\author{R. J. Cava$^2$}
\author{N. P. Ong$^{1,*}$}
\affiliation{
Departments of Physics$^1$ and Chemistry$^2$, Princeton University, Princeton, NJ 08544
} 

\date{\today}      
\begin{abstract}
{\bf 
Dirac semimetals and Weyl semimetals are 3D analogs of graphene in which crystalline symmetry protects the nodes against gap formation~\cite{Kane,Ashvin,Bernevig}. Na$_3$Bi and Cd$_3$As$_2$ were predicted to be Dirac semimetals~\cite{Wang1,Wang2}, and recently confirmed to be so by photoemission~\cite{Borisenko,Chen,Hasan}. Several novel transport properties in a magnetic field $\bf H$ have been proposed for Dirac semimetals~\cite{Son,Ashvin,Balents,Hosur}. Here we report an interesting property in Cd$_3$As$_2$ that was unpredicted, namely a remarkable protection mechanism that strongly suppresses back-scattering in zero $\bf H$. In single crystals, the protection results in ultrahigh mobility, $9\times 10^6$ cm$^2$/Vs at 5 K. Suppression of backscattering results in a transport lifetime 10$^4\times$ longer than the quantum lifetime. The lifting of this protection by $\bf H$ leads to a very large magnetoresistance. We discuss how this may relate to changes to the Fermi surface induced by $\bf H$. 
}
\end{abstract}
 
\maketitle      
In the 3-dimensional Dirac semimetal, the node at zero energy is protected against gap formation by crystalline symmetry~\cite{Kane,Ashvin,Bernevig}. Predictions~\cite{Wang1,Wang2} that Cd$_3$As$_2$ and Na$_3$Bi are Dirac semimetals have recently been confirmed by angle-resolved photoemission~\cite{Borisenko,Chen,Hasan}. When time-reversal symmetry (TRS) is broken, the Dirac semimetal is expected to evolve to a Weyl semimetal. This has stimulated intense interest in the possibility of observing ``charge-pumping'' effects in the Weyl state~\cite{Son,Ashvin,Balents,Hosur}. Here we report an unpredicted transport property. Below 5 K in zero magnetic field, Cd$_3$As$_2$ displays ultrahigh mobility ($9\times 10^6$ cm$^2$/Vs). The dramatic suppression of the high residual conductivity in a magnetic field $H$ implies that the carriers are protected against backscattering by an unknown mechanism.

\begin{figure*}[t]
\includegraphics[width=11 cm]{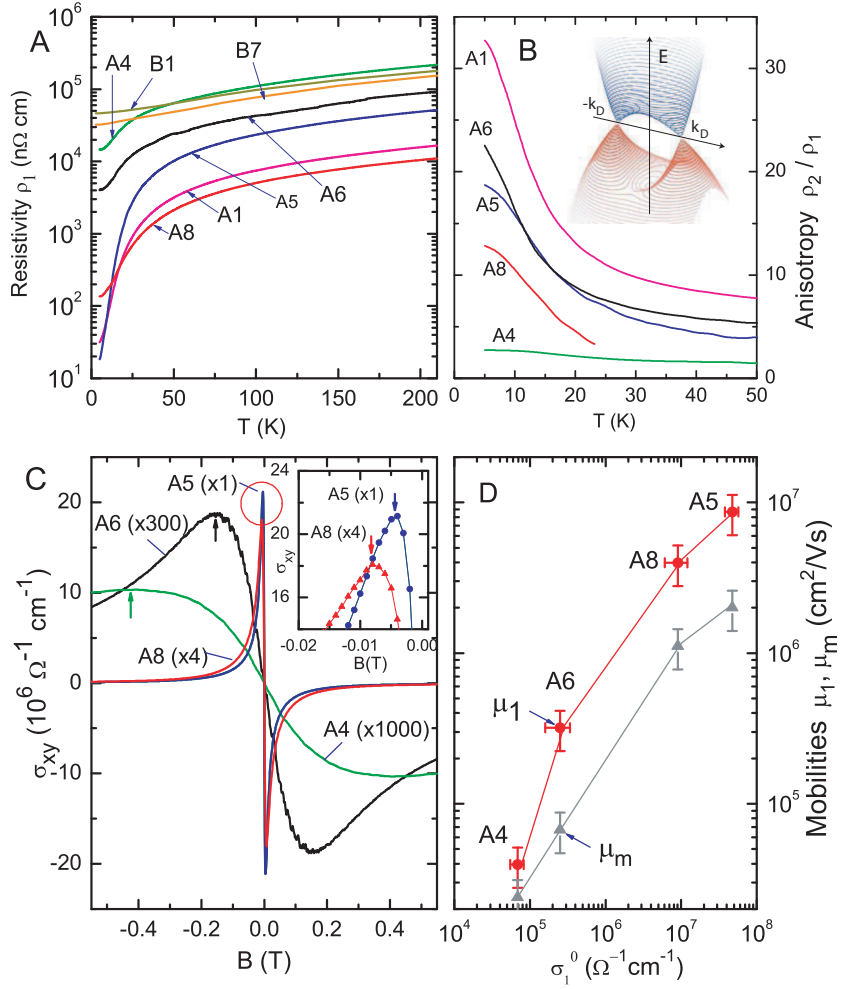}
\caption{\label{figRT} 
Transport measurements in a series of Cd$_3$As$_2$ samples. Panel A: Curves of the resistivity $\rho_1$ vs. $T$ measured along the needle axis $\bf\hat{x}$ in 5 Set A and 2 Set B samples (semilog scale). In needle-shaped crystals (Set A), $\rho_1$ undergoes a steep decrease below 50 K that is strongly sample dependent. In A5, $\rho_1$ falls by 3 orders of magnitude to 21 n$\Omega$cm at 5 K. In A4, however, $\rho_1$ has a milder decrease (to 14.6 $\mu\Omega$cm at 5 K). By contrast, the multidomain samples B1 and B7 do not display the steep decrease below 50 K. Panel B shows that the anisotropy $\gamma \equiv \rho_2/\rho_1$ at 5 K is large (20-30) in A1 and A5, but modest for A4 (2.7). The inset is a sketch of the energy dispersion $E(k)$ near the Dirac nodes (adapted from Ref. \cite{Yazdani}). Panel C plots the Hall conductivity $\sigma_{xy}$ vs. $B$ in A4, A5, A6 and A8 ($B=\mu_0 H$ with $\mu_0$ the vacuum permeability). The peak locates the geometric-mean mobility $\mu_m \equiv \sqrt{\mu_1\mu_2}$. For clarity, the region encircled by the red circle is shown expanded in the inset. Panel D plots the measured mobility $\mu_m$ (solid triangles) and the $x$-axis mobility $\mu_1 = \mu_m\sqrt\gamma$ vs. the zero-$H$ conductivity $\sigma^0_1$ for A4, A5, A6 and A8. 
}
\end{figure*}

Crystals of Cd$_3$As$_2$, grown by a flux technique (Supplementary Information SI) are needle-like with well-defined facets. The longest axis lies along ($1\bar{1}0$) and the largest face is normal to (112). In addition to these ``Set A'' samples, we also investigated multidomain samples which lack defined facets (Set B). Cd$_3$As$_2$ is unusual in that exactly $\frac14$ of the 64 Cd sites in each unit cell are vacant in the ideal lattice~\cite{Maz}. We have found that a rich spectrum of transport properties exists even among crystals extracted from the same boule. The residual resistivity and mobility (at 5 K) can vary by a factor of 200 (Table \ref{Tab1}). A remarkable pattern reflecting this variation is already apparent in Fig. \ref{figRT}A, which plots the $x$-axis resistivity $\rho_1$ vs. temperature $T$ (we take $\bf\hat{x}||$($1\bar{1}0$) and ${\bf\hat{z}}||(112)$; subscripts 1 and 2 refer to axes $x$ and $y$, respectively). Above 50 K, the resistivity profiles in Set A samples are similar. However, as $T$ decreases from 50 to 5 K, $\rho_1$ falls steeply, implying a strong enhancement in the transport lifetime $\tau_{tr}$. In A5, the enhancement results in a residual resistivity ratio RRR of 4,100 and a residual resistivity considerably lower than that in high-purity Bi~\cite{Lawson,Hartman} (21 vs. 100 n$\Omega$ cm). By contrast, this enhancement is completely absent in samples B1 and B7. A first clue to the enhancement in $\tau_{tr}$ comes from examining the resistivity anisotropy $\gamma(T) = \rho_2/\rho_1$. Using the Montgomery technique~\cite{Montgomery}, we have determined that $\gamma(T)$ increases monotonically with decreasing $T$. As shown in Fig. \ref{figRT}B, $\gamma(T)$ at 5 K rises to 20-30 in samples with large lifetime enhancements (A1 and A5), whereas $\gamma$ is only 2.7 in A4 which has the smallest enhancement (Table \ref{Tab1}). (To rule out the possibility that the very small $\rho_1$ results from a thin surface layer of Cd, we have carried out several tests described in the SI.) 

\begin{figure*}[t]
\includegraphics[width=12 cm]{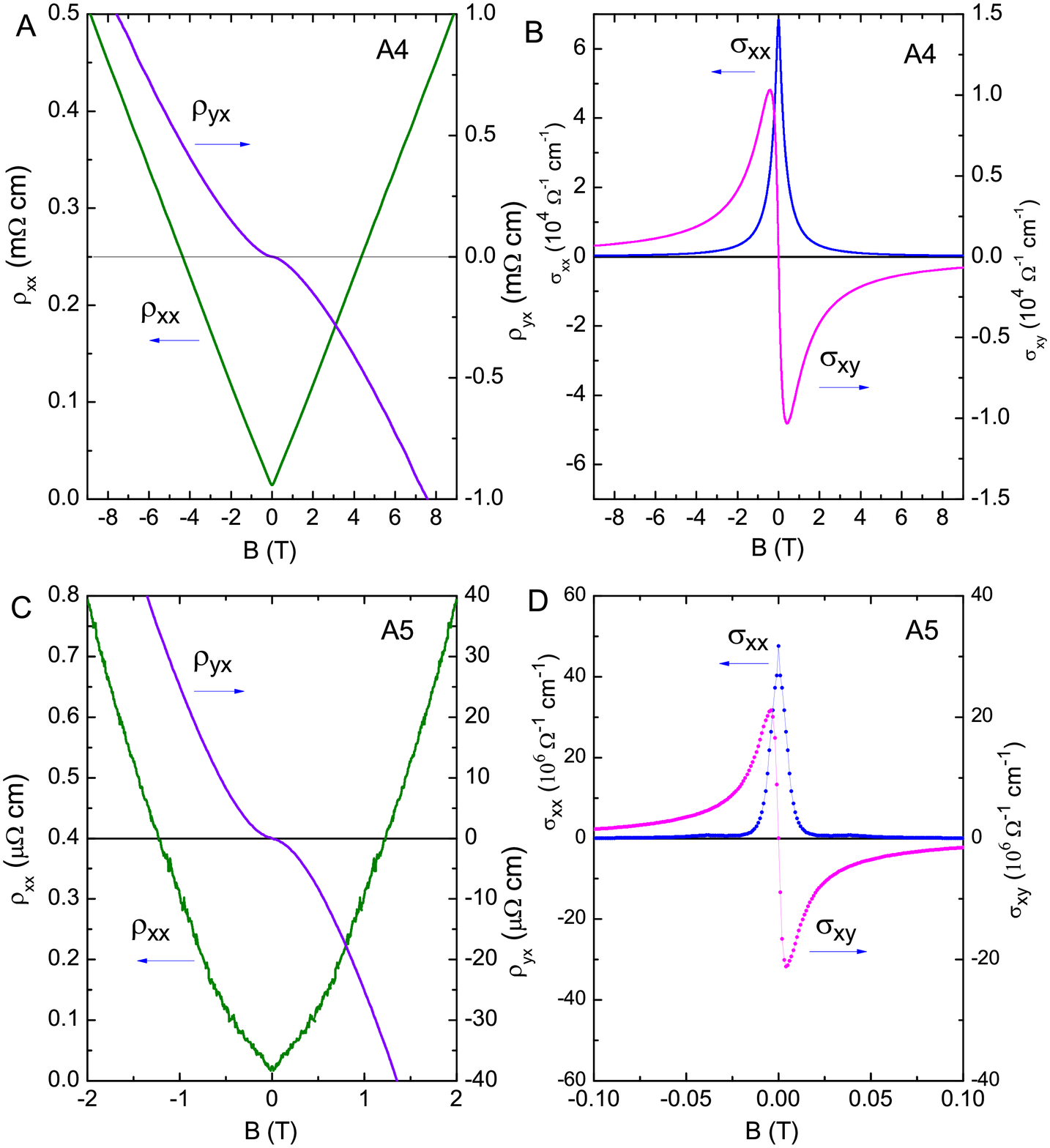}
\caption{\label{figMRCond} 
Conversion of resistivity matrix $\rho_{ij}$ to conductivity matrix $\sigma_{ij}$. In Sample A4 (Panel A), the resisitivty $\rho_{xx}$ displays an unusual $H$-linear profile while the Hall resistivity $\rho_{yx}$ ($n$-type in sign) has a weak anomaly in weak $H$ (measured at 5 K with $\bf H||\hat{z}$ and current $\bf I||\hat{x}$; $B=\mu_0 H$). The inferred conductivity $\sigma_{xx}(H)$ and Hall conductivity $\sigma_{xy}(H)$ are plotted in Panel B. The sharp extrema in $\sigma_{xy}$ at $\pm$ 0.42 T locate the geometric-mean mobility $\mu_m = \sqrt{\mu_1\mu_2}$. Panel C plots $\rho_{xx}$ and $\rho_{yx}$ at 5 K in Sample A5. The corresponding curves of $\sigma_{ij}(H)$ are in Panel D. Now the peaks in $\sigma_{xy}$ occur at $\pm$5.0 mT reflecting the much higher $\mu_m$ in A5 (by a factor of 85). In A5, the MR is also larger but becomes $H^2$ at large fields. Curves for samples A6 and A8 are shown in the SI. Typical dimensions of the crystals are $1.5\times 0.3\times 0.2$ mm$^3$ (see Table S1 in SI for exact dimensions). 
}
\end{figure*}

The results in Figs. \ref{figRT}A and B suggest that, at low $T$, the carrier mobilities $\mu_1$ and $\mu_2$ become very large but may be highly anisotropic. Employing the magnetic field as a ``yardstick'', we have managed to determine the mobility directly by measuring the resistivity tensor $\rho_{ij}(H)$ to high resolution in the weak-field regime. As discussed below (see Fig. \ref{figMRCond}), curves of $\sigma_{xy}(H)$ are obtained by inverting the matrix $\rho_{ij}$. In all samples, $\sigma_{xy}(H)$ exhibits the ``dispersive-resonance'' profile with sharp peaks that reflect the elliptical cyclotron orbit executed in weak $H$. In standard Bloch-Boltzmann transport, the reciprocal of the peak field $1/B_{max}$ equals the geometric mean of the mobilities $\mu_m\equiv \sqrt{\mu_1\mu_2}$. Hence, with $\gamma(T)$ known, we may obtain $\mu_1$ and $\mu_2$. (As a check, we have measured $B_{max}$ of $\sigma_{xy}$ at several $T$ in one sample (A5). As shown in SI, we find that $\mu_m(T)$ and $\mu_1(T)$ track the steep decrease in $\sigma^0_1$ as $T$ increases from 5 to 100 K.)

\begin{table}
\begin{tabular}{|c|c|c|c|c|c|c|} \hline
Sample	& $\rho_1$	       & $\gamma$	& RRR & $\mu_1$                & MR(9T)    &  $n_H$ (9T)         \\ \hline
(units)   & n$\Omega$cm &     --          &  --    &  cm$^2$/Vs            &      --       & $10^{18}$ cm$^{-3}$\\ \hline\hline
A1         &   32                 &   32.7        &  781  &   $\sim 3\times 10^6$* & 582  &   9.1           \\ \hline
A4         &   14,600         &   2.72	       &   21.4  &  40$\times 10^3$  &  34.5      &    4.4                 \\ \hline
A5         &   21               &   18.7       &  4,100  &  8.7$\times 10^6$  &  1,336      &   7.4             \\ \hline
A6        &    4,000          &   22.6       &  32.2    &   320 $\times 10^3$ &  112     &    12.0                \\ \hline
A8       &    110             &   12.8        &  118    &   4.0 $\times 10^6$  &  404     &      13.3       \\ \hline
B1      &    46,500        &    --     &   5.37   &   $\sim 10\times 10^3$* &  36.9  &   --       \\ \hline
B7      &    32,200       &    --      &  7.26   &   $\sim 20\times 10^3$*  &  62.2 &   15        \\ \hline
\end{tabular}
\caption{\label{Tab1}
Parameters of the 7 samples investigated. $\rho_1$ is the resistivity along $\bf\hat{x}$ at 5 K. The anisotropy $\gamma$ is $\rho_2/\rho_1$ at 5 K ($\gamma$ is undefined in B1 and B7). RRR is the ratio $\rho_1(300)/\rho_1(5)$.  The mobilities are determined from $\sigma_{xy}$ and $\gamma$, except in A1, B1 and B7 (*) where they are estimated from the residual resistivity. MR is the ratio $\rho_{xx}(9T)/\rho_{xx}(0)$ at 5 K. The Hall density $n_H$(9T) equals $B/e\rho_{yx}$ measured at 9 T (all $n$-type). 
}
\end{table}

\begin{figure*}[t]
\includegraphics[width=15 cm]{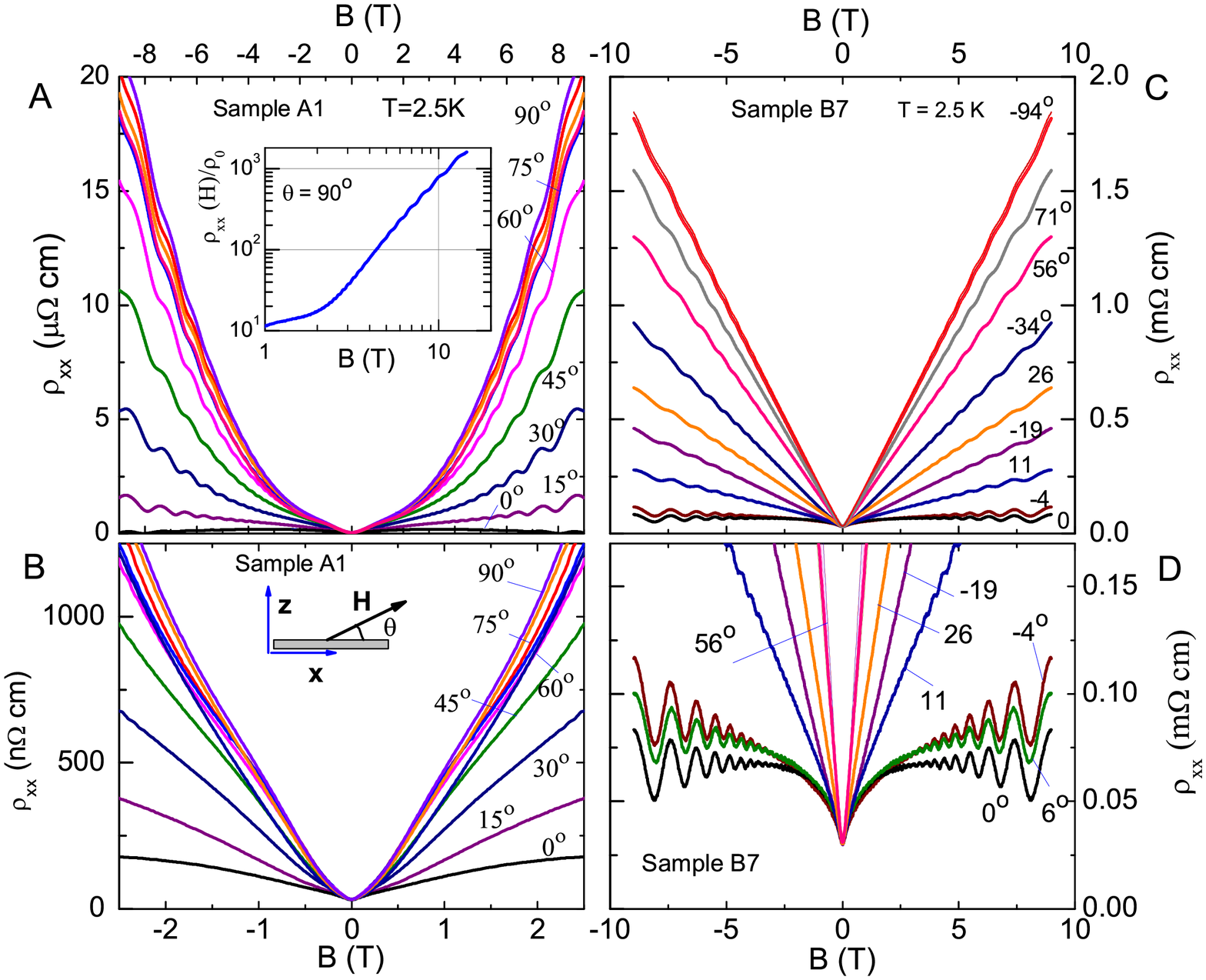}
\caption{\label{figMR} 
Magnetoresistance curves $\rho_{xx}(H,\theta)$ and SdH oscillations in tilted $\bf H$ in Cd$_3$As$_2$ at 2.5 K in Samples A1 and B7. In Panel A, MR curves for the high-mobility single crystal A1 are plotted for $0<\theta<90^{\circ}$. The log-log plot in the inset shows that, at 2 T, $\rho_{xx}(H)/\rho_{0}$ changes from an $H$-linear increase to an anomalous power-law $H^{2.55}$, reaching a value of 1,600 at 15 T ($\theta = 90^{\circ}$). SdH oscillations are resolved at all $\theta$. In weak $H$ (Panel B), the MR is nearly $H$-linear. As $\theta\to 0$, the MR rapidly decreases (at fixed $H$). It acquires a negative contribution for $|\theta|< 5^\circ$. The multidomain sample B7 displays a similar behavior except that the striking $H$-linear dependence persists to 9 T (Panels C and D). The tilt angle $\theta$ and the $x$- and $z$-axes are defined in the inset in Panel B [$\bf\hat{z}||$(112)]. 
}
\end{figure*}

As shown in Fig. \ref{figRT}C, the curves of $\sigma_{xy}(H)$ at 5 K in A4, A5, A6 and A8 display the dispersion profile described. Remarkably, $B_{max}$ shrinks by a factor of 85 (420 mT to 5 mT) as $\mu_m$ increases across the samples. The large variation in $\mu_1$ and $\mu_m$ scales well with the residual conductivity $\sigma^0_1$ (Fig. \ref{figRT}D). Hence we conclude that the anomalously low residual resistivities arise from mobilities that attain ultrahigh values of 10$^7$ cm$^2$/Vs, far higher than in previous studies~\cite{Rosenman,Iwami,Neve}. For comparison, the highest electron mobility in Bi is reported~\cite{Hartman} to be $9\times10^7$ cm$^2$/Vs (see SI). The highest mobility observed to date in the 2D electron gas in an AlGaAs/GaAs heterojunction is $3.6\times 10^7$ cm$^2$/Vs \cite{Pfeiffer}. [Despite the 100-fold change in $B_{max}$, the curves of $\sigma_{xy}(H)$ in the 4 samples collapse to the same curve when plotted in scaled variables (Fig. S6 of SI). In the SI (Sec. S3), we describe how the scaling excludes the scenario of a highly disordered system with a broad distribution of lifetimes.]

We turn next to the giant MR observed in all samples. Figure \ref{figMRCond} shows the curves of $\rho_{ij}(H)$ in A4 and A5, along with curves of $\sigma_{ij}(H)= [\rho_{ij}]^{-1}$ obtained by matrix inversion (similar plots for A6 and A8 are in SI). In transverse field ($\bf H||\hat{z}$), the needle crystal with the lowest mobility A4 ($\mu_1$= 4.0$\times 10^4$ cm$^2$/Vs) shows a striking $H$-linear MR profile (Fig. \ref{figMRCond}A). All Set B samples also display the $H$-linear MR (see SI). From the Hall resistivity $\rho_{yx}$ at large $H$, we obtain an $n$-type ``Hall density'' $n_H= B/e\rho_{yx} \sim 4.4\times 10^{18}$ cm$^{-3}$ at 9 T (Table \ref{Tab1}). In A5, with the highest $\mu_1$ (Panel C), the MR is significantly larger, but now has the form $H^\alpha$ with $\alpha$ = 2--2.5 above $\sim$2 T (the trend from $H$-linear to $H^\alpha$ with increasing $\mu_1$ is robust). 

Measurements of the MR and Shubnikov de Haas (SdH) oscillations in a tilted $\bf H$ provide further insight on the enhanced lifetime (we fix $\bf H$ in the $x$-$z$ plane at an angle $\theta$ to $\bf\hat{x}$). The MR in A1 at 2.5 K is displayed in Figs. \ref{figMR}A,B for several tilt angles $\theta$. (The MR ratio is defined as $\rho_{xx}(T,H)/\rho_1(T,0)$;  see Table \ref{Tab1}). A log-log plot of the MR in A1 is plotted in the inset of Fig. \ref{figMR}A. As $\bf H$ is tilted towards $\bf\hat{x}$ ($\theta\to 0$), the MR decreases rapidly. In Fig. \ref{figMR}C, D we display the MR in Sample B7, which has a similar variation vs. $\theta$.

\begin{figure*}[t]
\includegraphics[width=16 cm]{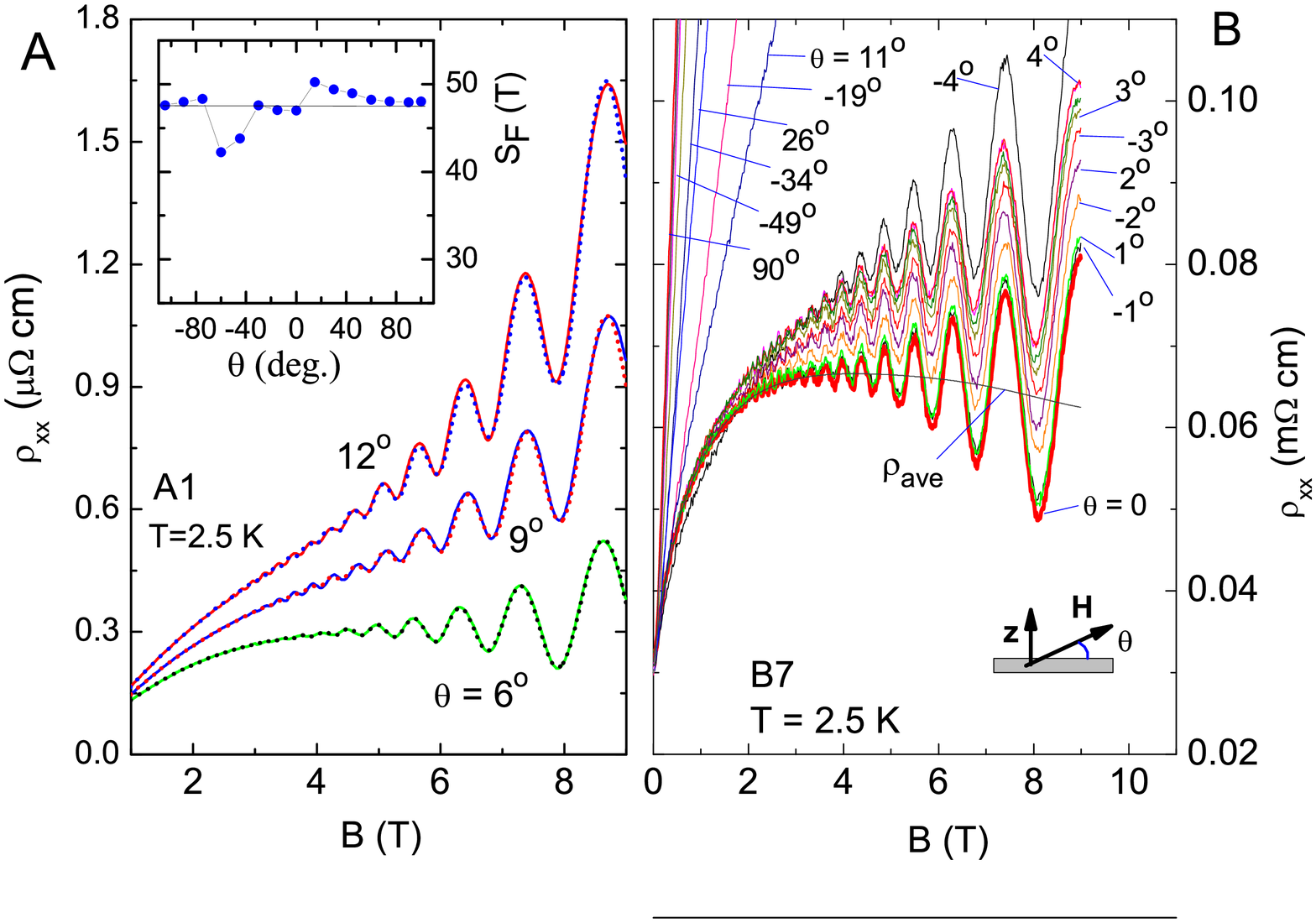}
\caption{\label{figSdH} 
Shubnikov de Haas (SdH) oscillations in Samples A1 and B7 at 2.5 K. Panel A shows the SdH oscillations in $\rho_{xx}$ (solid curves), for tilt angles $\theta$ = 6$^\circ$, 9$^\circ$ and 12$^\circ$. The fits to the Lifshitz-Kosevich expression (shown as dotted curves) yield a quantum lifetime $\tau_Q$ $\sim 10^4\times$ shorter than $\tau_{tr}$. The inset plots the variation of $S_F(\theta)$ about the average 47.5 T. Panel (B) shows traces of $\rho_{xx}$ vs. $H$ as $\theta$ is changed by $1^{\circ}$ steps through 0$^\circ$ at 2.5 K (Sample B7). The curve at $\theta=0$ (bold) has a distinct, negative MR contribution (shown by the averaged plot $\rho_{ave}$). As $|\theta|$ increases from $1^\circ \to 4^\circ$, the giant positive MR term rapidly dominates. Below 0.2 T, the MR is positive and nearly isotropic (see SI for more information).
}
\end{figure*}

To highlight the SdH oscillations, we plot traces of $\rho_{xx}$ in A1 for $\theta$ = 6$^\circ$, 9$^\circ$ and 12$^\circ$ in Fig. \ref{figSdH}A. In sharp contrast to the MR, varying $\theta$ has very little effect on the cross-section $S_F$ of the Fermi Surface (FS) inferred from the SdH period in all samples. The weak variation of $S_F$ with $\theta$ (inset) implies a nearly spherical FS and isotropic $v_F$, in good agreement with earlier experiments~\cite{Rosenman,Neve}. This contrasts with the strong anisotropy $\gamma$ shown in Fig. \ref{figRT}B (see below). Band calculations~\cite{Wang2} and a recent STM study~\cite{Yazdani} reveal Dirac nodes at $(0,0,\pm k_D)$ with $k_D\sim 0.032\;\mathrm{\AA}^{-1}$, and the Fermi energy $E_F$ lying in the conduction band (inset, Fig. \ref{figRT}B). At each $\theta$, the SdH oscillations in both A1 and B7 fit very closely the Lifshitz-Kosevich expression with a single frequency (see Fig. \ref{figSdH}A and SI). In addition to $S_F$, the fits yield a high Fermi velocity $v_F \, = \,9.3\times 10^5$ m/s, and a Fermi energy $E_F$ = 232 mV, consistent with recent STM~\cite{Yazdani} and ARPES experiments~\cite{Chen,Hasan,Borisenko}. The electron density $n = k_F^3/3\pi^2\simeq 1.9\times 10^{18}$ cm$^{-3}$ is a factor of 2-10 smaller than $n_H$ (Table \ref{Tab1}). Tracking the SdH signal to 45 T, we reach the $N$=1 Landau level (LL) at 27 T and begin accessing the $N$=0 LL above 36 T (see Fig. S8 of SI). As discussed in Sec. S4 of SI, the presence of a second band can be excluded to a resolution of 3$\%$ of the main SdH amplitude. Surprisingly, the quantum lifetime is found to be very short ($\tau_Q\, =\, 3-8.6\times 10^{-14}$ s), compared with $\tau_{tr}$ derived from $\mu_1$.

For a band with Dirac dispersion, the mobility is expressed as $\mu = ev_F\tau_{tr}/\hbar k_F$. Using $k_F$ and $\mu_1$ (Fig. \ref{figRT}D), we estimate $\tau_{tr}\sim 2.1\times 10^{-10}$ s in A5, corresponding to a ``transport'' mean-free-path $\ell_{tr}\sim$ 200 $\mu$m. Defining $R_\tau\equiv\tau_{tr}/\tau_Q$, we find that $R_\tau$ attains values 10$^4$ at 2.5 K. The large $R_{\tau}$ provides an important insight into the anomalously low resistivity. $\tau_{tr}$ measures (2$k_F$) back-scattering processes that relax current, whereas $\tau_Q$ is sensitive to all processes that cause Landau level (LL) broadening, including forward scattering. Hence $R_{\tau}$ generally exceeds 1. Still, $R_{\tau}$ here is exceptionally large compared with values (10-100) reported for GaAs-based 2DEG~\cite{Paalanen,Harrang,Coleridge}.

The picture that emerges is that, in zero field, there exists a novel mechanism that strongly protects the carriers moving parallel to $\bf\hat{x}$ against back-scattering, despite lattice disorder. In the case of the 2DEG in GaAs/AlGaAs, the large $R_{\tau}$ arises because the dopants are confined to a $\delta$-layer set back from the 2DEG~\cite{Paalanen,Harrang,Coleridge}. Charge fluctuations in the dopant layer lead only to small-angle scatterings, which strongly limit $\tau_Q$ but hardly affect $\tau_{tr}$. Here there is no obvious separation of the scattering centers from the conduction electrons, yet $R_{\tau}$ is even larger. As evident in Figs. \ref{figRT} and \ref{figMR}, the protection exists in zero $H$, but is rapidly removed by field. Since the FS is nearly isotropic in Cd$_3$As$_2$, the full anisotropy $\gamma$ comes from an anisotropic transport scattering rate $\Gamma_{tr} = 1/\tau_{tr}$. Moreover, as the anisotropy is rapidly suppressed above $\sim$ 20 K (Fig. \ref{figRT}B), the protection extends only to elastic scattering. It is interesting to contrast our results with ballistic propagation in carbon nanotubes. In nanotubes, the carriers can propagate between contact reservoirs without suffering any elastic collision. In our samples A1 and A5, the Dirac electrons at 5 K undergo a great number of collisions (predominantly forward scattering) which lead to severe broadening of the LL; but it takes 10$^4$ collisions to reverse the momentum. Hence $\ell_{tr}\gg \ell_0$ (the mfp between collisions). 

The giant MR is universally observed in all samples. We find the striking $H$-linear MR observed in the low-mobility samples (A4 and all Set B samples) especially interesting. Non-saturating $H$-linear MR is rare in metals and semimetals. It has been reported in Ag$_{2+\delta}$Se ($\delta\sim0.01$)~\cite{Rosenbaum,Zhang} and Bi$_2$Te$_3$~\cite{Qu}, both topological insulators. Abrikosov has derived an $H$-linear MR for Dirac electrons occupying the lowest LL~\cite{Abrikosov}. However, the $H$-linear MR here already exists at very low $H$. We remark on two notable features of the MR in B7. In the limit $H\to 0$, the MR becomes nearly isotropic (Figs. \ref{figMR}D and \ref{figSdH}B). This implies that a Zeeman coupling to the spin degrees is important (the g-factor is known to exceed 20). Further, when $T$ is raised to 300 K, the $H$-linear profile is unchanged, except that the cusp at $H$=0 becomes progressively rounded by thermal broadening. This robustness suggests that an unconventional mechanism for the $H$-linear MR. Both points are discussed further in the SI.

Our finding of a strongly $H$-dependent $\Gamma_{tr}$ is consistent with field-induced changes to the FS. In Dirac semimetals, breaking of TRS by $H$ rearranges the Dirac FS~\cite{Kane,Wang1,Ashvin,Balents,Hosur}. The FS either splits into two disjoint Weyl pockets (if $H$ couples to both spin and orbital degrees) or becomes two concentric spheres (if $H$ couples to spin alone)~\cite{Wang1}. Because these changes are linear in $H$, it would be interesting to see if they can lead to lifting of the protection mechanism and the giant MR observed.

In Dirac semimetals, there is strong interest in whether the chiral term $(e^3/4\pi^2\hbar^2){\bf E\cdot H}$ can be detected as a negative contribution to the longitudinal MR ($\bf E||H$), with $\bf E$ the electric field (see SI)~\cite{Hosur,Balents,Son,Nielsen,Sid}. Clearly, the giant positive MR has to be carefully considered since it constitutes a $\theta$-dependent ``background'' that is much larger than the chiral term (we estimate that, at 1 T, the latter decreases $\rho_{xx}$ by roughly $10^{-2}$). Although this seems daunting, we note that the competing terms are of opposite signs and are out-of-phase: the chiral term varies as $-\cos\theta$, whereas the positive MR term varies as $\sin\theta$ (vanishes at $\theta=0$). In Fig. \ref{figSdH}B, we plot the MR curves in B7 stepping $\theta$ in $1^\circ$ steps through 0$^\circ$. Clearly, $\rho_{xx}$ attains a sharp minimum which we identify as $\theta=0$ (bold curve), but swings up when $|\theta|$ exceeds $2^\circ$. In the curve at $\theta=0$, we resolve a weak, but distinct negative MR term (see the averaged curve $\rho_{ave}$). To compare this negative term with the chiral term in a physically meaningful way, we will need to apply larger $H$ and finer control of $\theta$. These experiments are being pursued. After completion of these experiments, we learned of the results in Refs. \cite{Fudan,IOP}.

{We thank Andrei Bernevig, Sid Parameswaran, Ashvin Vishwanath and Ali Yazdani for valuable discussions, and Nan Yao for assistance with EDX measurements. N.P.O. is supported by 
the Army Research Office (ARO W911NF-11-1-0379). R.J.C. and N.P.O. are supported by a MURI grant on Topological Insulators (ARO W911NF-12-1-0461) and by the US National Science Foundation
(grant number DMR 0819860). T.L acknowledges scholarship support from the Japan Student Services Organization. Some of the experiments were performed at the National High Magnetic Field Laboratory, which is supported by National Science Foundation Cooperative Agreement No. DMR-1157490, the State of Florida, and the U.S. Department of Energy.}

\newpage


\newpage
%
%

%
\renewcommand{\thefigure}{S\arabic{figure}}
\renewcommand{\thesection}{S\arabic{section}}
\renewcommand{\theequation}{S\arabic{equation}}
\renewcommand{\thetable}{S\arabic{table}}

\setcounter{equation}{0}
\setcounter{figure}{0}
\setcounter{table}{0}

\vspace{4mm}
{\bf Supplementary Information\\
} 
\vspace{6mm}

\section{Crystal growth, EDX spectra and X-ray diffraction}
Cd$_3$As$_2$ crystals were grown using excess Cd as a flux, with the overall ratio of Cd$_8$As$_2$. The elements were handled in a glovebox under an Argon atmosphere and sealed in an evacuated quartz ampoule with a quartz wool plug. The sample was heated at 800$^\circ$ C for 2 days and then cooled at 6 degrees per hour to 400$^\circ$ C, then held for two more days. The samples were centrifuged, and then reheated to 400$^\circ$ C and centrifuged a second time to remove excess Cd. Both needle-like (Set A) and large chunky crystals (Set B) were isolated from the resulting material.

\begin{figure}[h]
\includegraphics[width=8 cm]{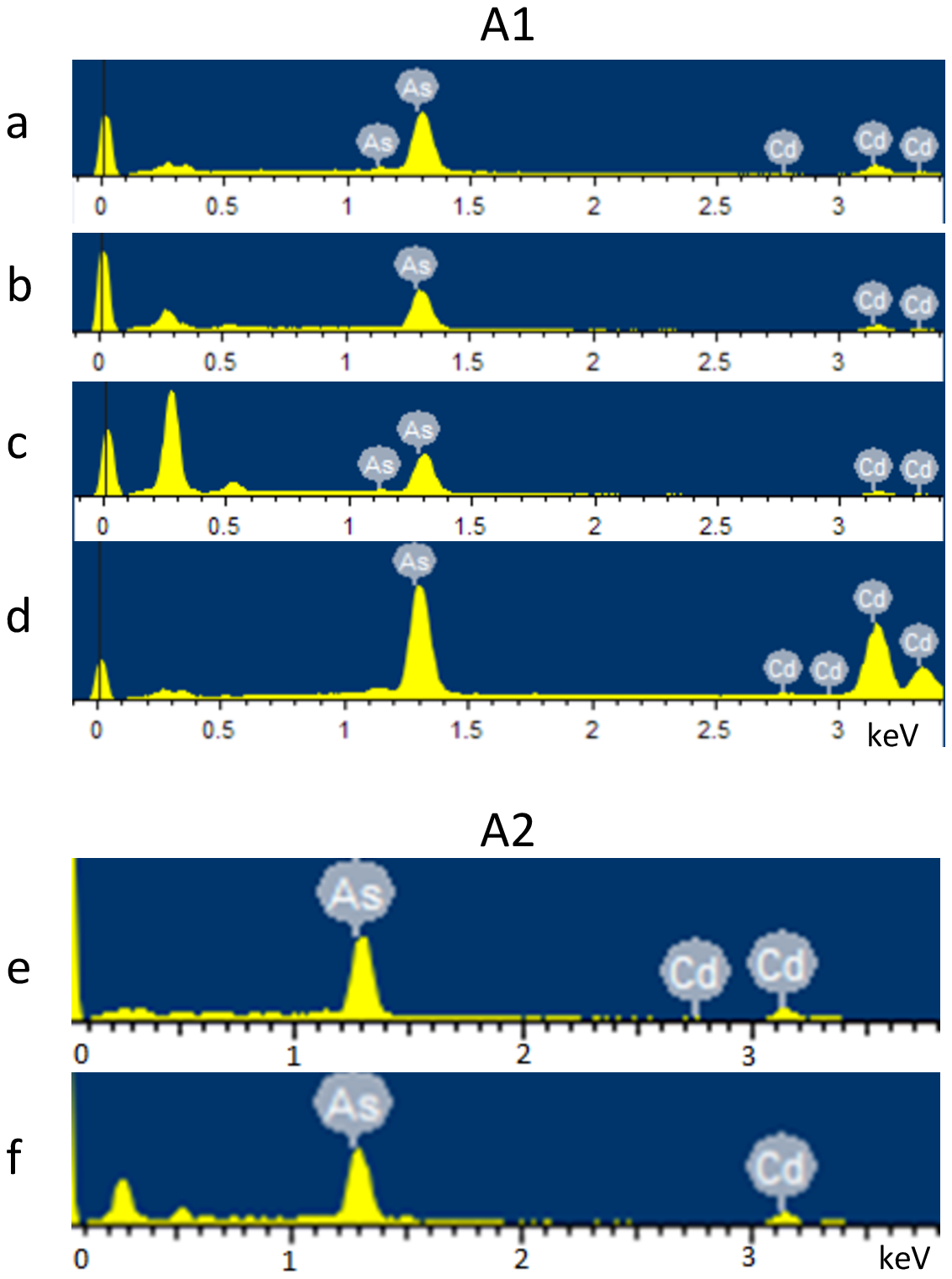}
\caption{\label{figEDX} 
EDX spectra for Samples A1 (Panels a, b, c, d) and A2 (Panels e and f). The energy of the incident beam is 5~keV (in Panels a, b, c, e, f) and 10~keV (Panel d). We define the beam's angle of incidence as $\phi$. In Panels a, d, e, f, $\phi = 0^\circ$ (normal incidence). In Panel b, $\phi = 45^\circ$. In Panel c, $\phi = 75^\circ$. The atomic percentages of As and Cd (As:Cd) in the individual panels are as follows. (a): 36.55 \%: 63.45 \%, (b): 39.74 \%: 60.26 \%, (c): 47.27 \%: 52.73 \%, (d): 34.36 \%: 65.64 \%, (e): 39.46 \%: 60.54 \%, (f): 40.44 \%: 59.56 \% (the ideal stoichiometric ratio is 40 \% : 60 \%). Within the uncertainties, the observed spectra shift to an As-rich composition as $\phi$ increases from 0$\to 75^\circ$ (a$\to$b$\to$c). 
}
\end{figure}

To eliminate the possibility that the very low residual resistivity $\rho_0$ observed in Set A crystals of Cd$_3$As$_2$ is due to the presence of a thin layer of elemental Cd on the crystal surface, Energy Dispersive X-ray Spectroscopy (EDX) analysis and Scanning Electron Microcopy (SEM) images were taken on a FEI Quanta 200 FEG Environmental SEM system. In Fig. \ref{figEDX}, we show a small subset of the EDX spectra obtained in Samples A1 (Panels a-d) and A2 (e and f). In order to probe the surface composition, multiple spots on the high mobility crystal described here were sampled with both a 10 keV and 5 keV incident beam as well as with a 5 keV beam at the two angles of incidence, $\phi$ = 45$^\circ$ and 75$^\circ$. No evidence of any surface layer of Cd was observed. Using the Kanaya-Okayama formula~\cite{Kanaya} for penetration depth, the penetration depth of a 5 keV beam in pure Cd is about 150 nm. [Using the published $\rho_0$ of elemental Cd, 0.1-1 n$\Omega$ cm, we calculate that the Cd film has to have an average thickness $t>$ 300 nm (40 nm) to mimic the observed $\rho_0$ in Sample A1 (A2).] Under these conditions, any layer of Cd would have been observed, at least, as a deviation towards a Cd-rich stoichiometry either upon lowering the beam energy or increasing the angle of incidence (measured relative to the normal). This was not observed. In fact, a deviation towards an As-richer stoichiometry was consistently observed at lower incident energies and higher incident angle $\phi$. Further, no features in the SEM images of either the surface or cross section of the crystal suggested any Cd layers or inclusions.

\begin{figure}[h]
\includegraphics[width=9 cm]{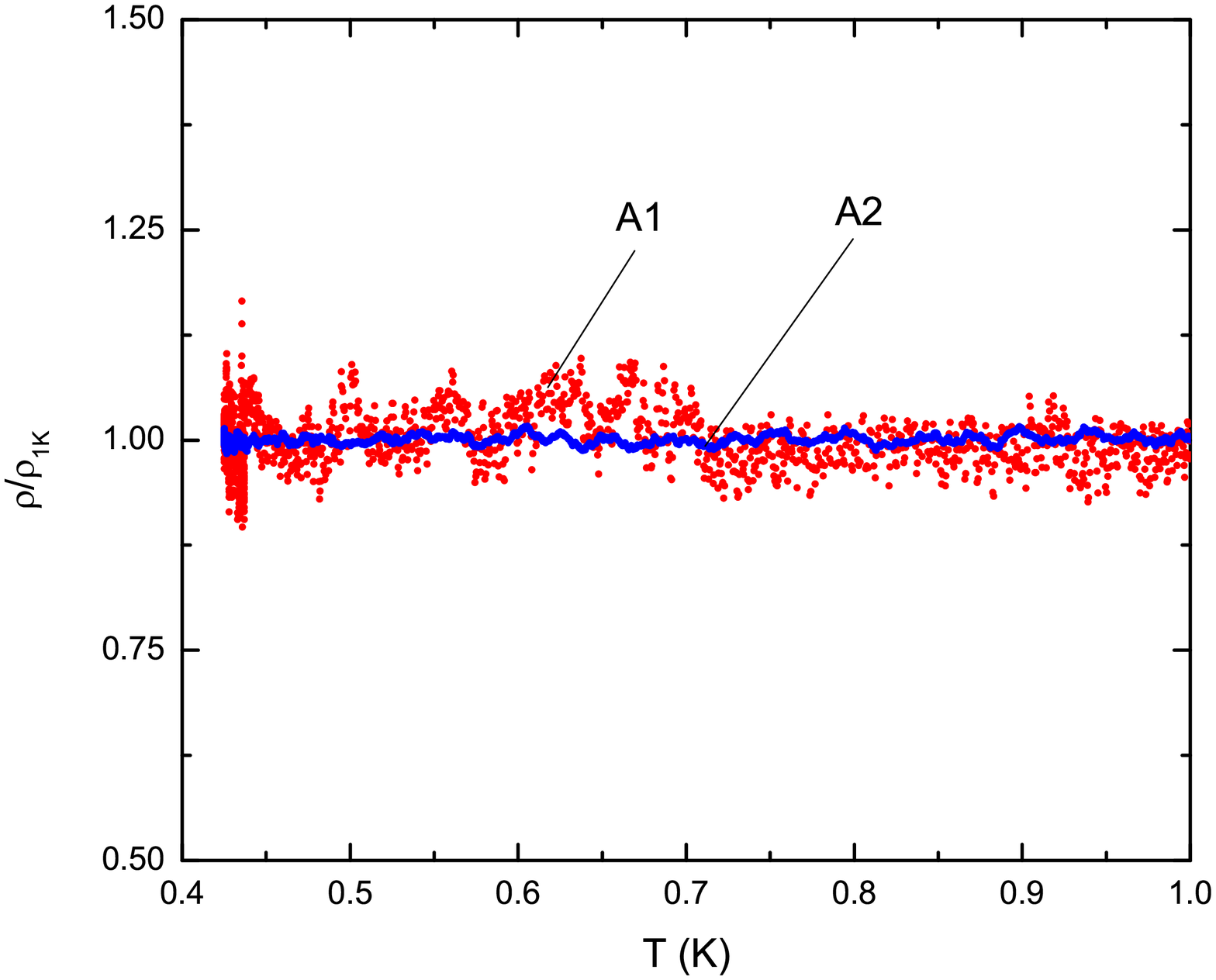}
\caption{\label{SC_check}
Resistivity normalized to values at 1 K in Samples A1 (red circles) and A2 (blue circles) plotted vs. $T$ between 0.4 and 1.0 K. No signatures of bulk superconductivity or fluctuation superconductivity were observed in this temperature interval (and higher). The absence strongly argues against the existence of a thin film of Cd plating the crystals (or a Cd spine extending through the bulk). The critical temperature of elemental Cd is 0.56 K. 
}
\end{figure}

\begin{figure}[t]
\includegraphics[width=8 cm]{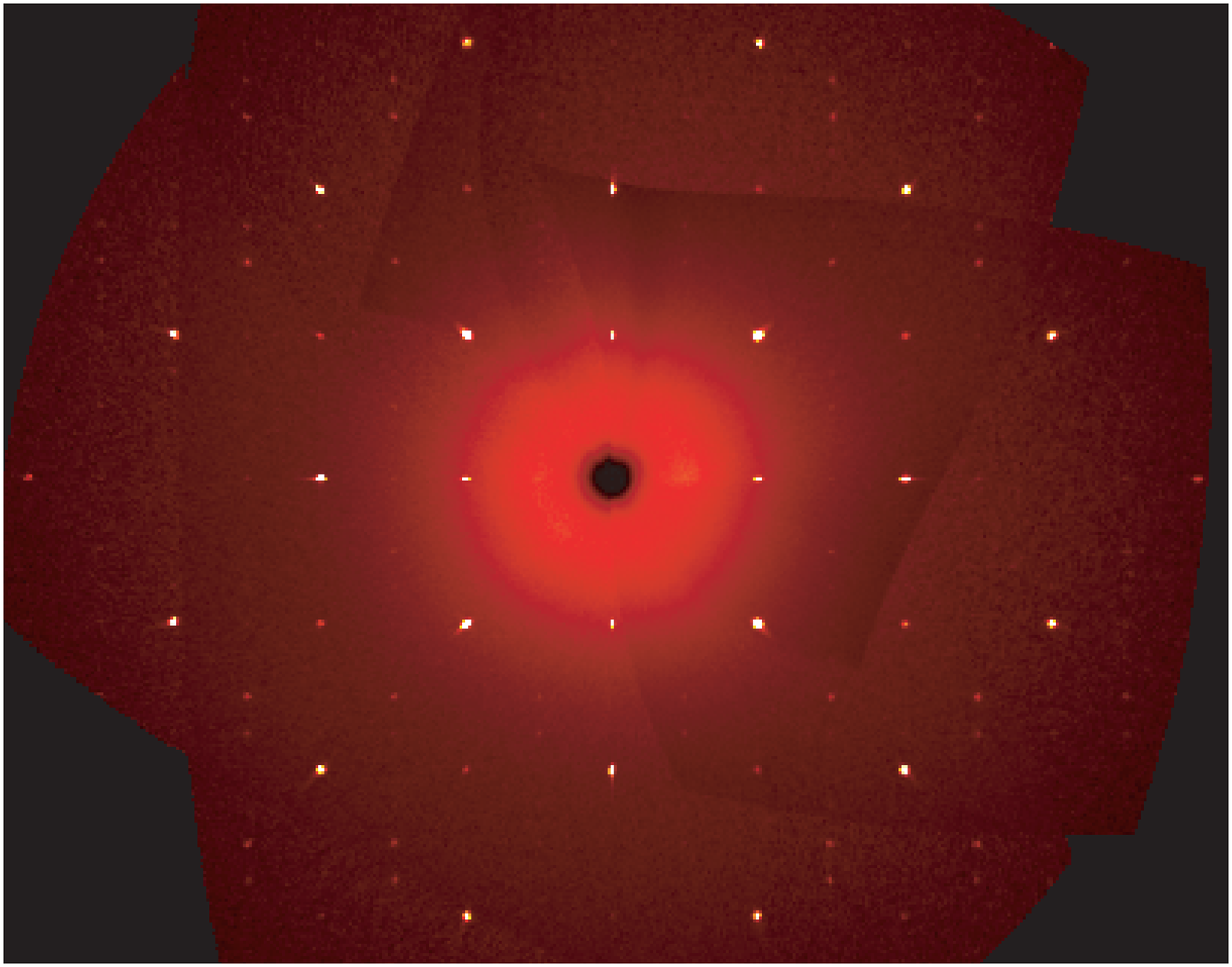}
\caption{\label{Xray}
Single-crystal X-ray diffraction precession image of the 0kl plane in the reciprocal lattice of Cd$_3$As$_2$ obtained on a segment (0.04 mm $\times$ 0.04 mm $\times$ 0.4 mm in size) of Sample A1. The weaker supercell reflections, which argue for the larger tetragonal cell, may be seen in between the bright spots. No diffuse scattering is seen.  All the resolved spots fit the crystal lattice structure recently established for Cd$_3$As$_2$ (Ref. ~\cite{Maz}).
}
\end{figure}

In addition, we searched for transport signatures of superconductivity in Set A samples ($T_c$= 0.56 K in elemental Cd) in $H=0$. 
As shown in Fig. \ref{SC_check}, no evidence for bulk or fluctuation superconductivity was observed in 2 samples (A1 and A2) measured down to 0.4 K. This strongly precludes either a thin surface Cd film or a bulk inclusion that extends over a significant segment of the crystal. Finally, measurements of the SdH were taken to 45 T at 0.3 K~\cite{Liang}. No evidence for additional SdH peaks was found (apart from the ones associated with the small electron pocket, as described).

Single-crystal X-ray diffraction (SXRD) was performed on a 0.04 mm $\times$ 0.04 mm $\times$ 0.4 mm crystal on a Bruker APEX II diffractometer using Mo K-alpha radiation (lambda = 0.71073 A) at 100 K. Exposure time was 35 seconds with a detector distance of 60 mm. Unit cell refinement and data integration were performed with Bruker APEX2 software. A total of 1464 frames were collected over a total exposure time of 14.5 hours. 21702 diffracted peak observations were made, yielding 1264 unique observed reflections collected over a full sphere. The crystal structure was refined using the full-matrix least-squares method on $F^2$, using SHELXL2013 implemented through WinGX. An absorption correction was applied using the analytical method of De Meulenaer and Tompa implemented through the Bruker APEX II software.

While the detailed SXRD measurements for structural refinement were carried out on small crystals, for physical property measurements, larger crystals are employed. The rod characterized in this study with dimensions of ~0.15 mm by ~0.05 mm by 1.2 mm, was used in order to ascertain the growth direction of the needle (long axis). The crystal was mounted onto a flat kapton holder and the Bruker APEX II software was used to indicate the face normals of the crystal after the unit cell and orientation matrix were determined. The long axis of the needle was found to be the [$1\bar{1}0$] direction. After the MR experiments were completed, a fragment of Sample A1 was also investigated by SXRD measurements and confirmed to also have its needle axis along [$1\bar{1}0$] (Fig. \ref{Xray}).

In our experience, the very high conductivity observed below 10 K in Set A samples degrades (albeit very slowly) when the crystals are stored at room temperature but exposed to ambient atmosphere. The degradation could arise from surface oxidation or gradual changes away from stoichiometry in the composition. Measurements of $\rho_0$ in A1 performed at NHMFL 3 months after our in-house experiments revealed that the zero-field residual resistivity $\rho_0$ had increased from 32 n$\Omega$ cm to 110 n$\Omega$ cm. This aging results in a factor of ~3.45 difference in the MR ratio $\rho(H)/\rho_0$ measured at 1 T in the high-field and low-field polar plots shown in Fig. \ref{figPolar10}.

\section{Measurement of anisotropy using Montgomerty Method}
The Montgomery method was used to determine the anisotropy $\rho_2/\rho_1$($\equiv \rho_y/\rho_x$)~\cite{Montgomery}. In this method, the anisotropic solid is identified with its isotropic ``equivalent''.

By scaling arguments, the ratio of the sample's resistivities $\rho_x$ and $\rho_y$ is related to the isotropic equivalent's dimensions by~\cite{Montgomery} 
\be
(\rho_y/\rho_x)^{1/2} = l_y/l_x \times l/w,
\label{eq:rho}
\ee
where $l_x$ and $l_y$ are the (unknown) lengths along $x$ and $y$ axes of the isotropic equivalent, and $l$ and $w$ are the known lengths of the original anisotropic sample measured along its $x$ and $y$ axes. We need $l_y/l_x$ to determine the anisotropy.

\begin{figure}[t]
\includegraphics[width=9 cm]{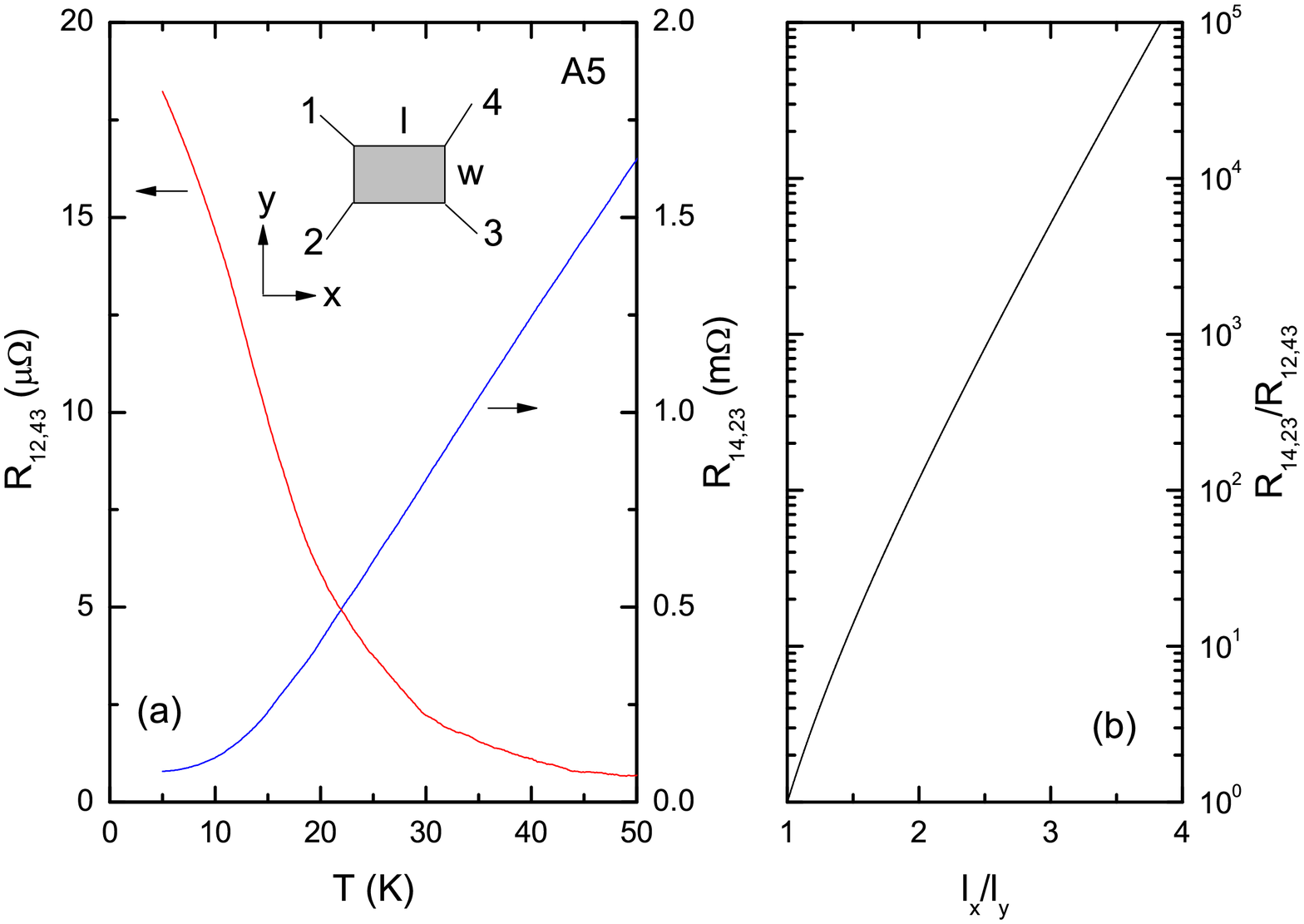}
\caption{\label{figMont} 
Montgomery method for determining the anisotropy. Panel (a) plots the measured nonlocal resistances $R_{12,43}$ and $R_{14,23}$ versus $T$ in sample A5. In Panel (b), we display the function that maps the ratio $R_{14,23}/R_{12,43}$ to the ratio $l_x/l_y$.
}
\end{figure}

It turns out that the ratio $l_y/l_x$ is uniquely determined by measuring two nonlocal resistances.
Four contacts (1,2,3,4) were attached along the four edges of the sample, indexed in cyclical order. At each temperature $T$, the nonlocal resistances $R_{12,43}$ ($\equiv$ V$_{43}$/I$_{12}$) and $R_{14,23}$ ($\equiv$ V$_{23}$/I$_{14}$) were measured. As an example, the measured $R_{12,43}$ and $R_{14,23}$ in sample A5 are plotted versus $T$ in Fig.~\ref{figMont}(a).

The ratio $R_{14,23}/R_{12,43}$ is uniquely mapped to $l_y/l_x$ using a function calculated by Logan, Rice and Wick~\cite{Logan} (the function is plotted in Fig.~\ref{figMont}(b)). Using the measured $R_{14,23}/R_{12,43}$, we may then find $l_y/l_x$ at each $T$. Finally, from Eq. \ref{eq:rho}, we calculate the anisotropy $(\rho_y/\rho_x)$.

\section{Sample Parameters}
Table \ref{TabS1} reports the physical dimensions of the 5 Set A crystals used in the experiment. 

\begin{table}[h]
\begin{tabular}{|c|c|c|c|c|} \hline
Sample	& $l_c$	   & $w$	      &  $t$       & $l_{tot}$ 	\\ \hline
units	      & mm	      & mm	      & mm	      & mm	     \\ \hline\hline
A1	      & 1.1	     &	0.2	       &	 0.1	      &	1.87			\\ \hline
A4	      &	2.1	      &	0.81	      &	0.65	    &	2.9				\\ \hline
A5	      &	0.9	      &	0.32	      &	0.35	    &	1.75				\\ \hline
A6	      &	0.55	    &	0.25	      &	0.21	    &	1.2				\\ \hline
A8	      &	0.5	      &	0.2		      & 0.1	      &	1.15				\\ \hline
\end{tabular}
\caption{\label{TabS1}
The dimensions (in mm) of the 5 Set A crystals investigated. $l_{tot}$ and $l_c$ are the total length and the distance between voltage contacts, respectively. $w$ and $t$ are the width and thickness, respectively.
}
\end{table}

\begin{figure}[t]
\includegraphics[width=9 cm]{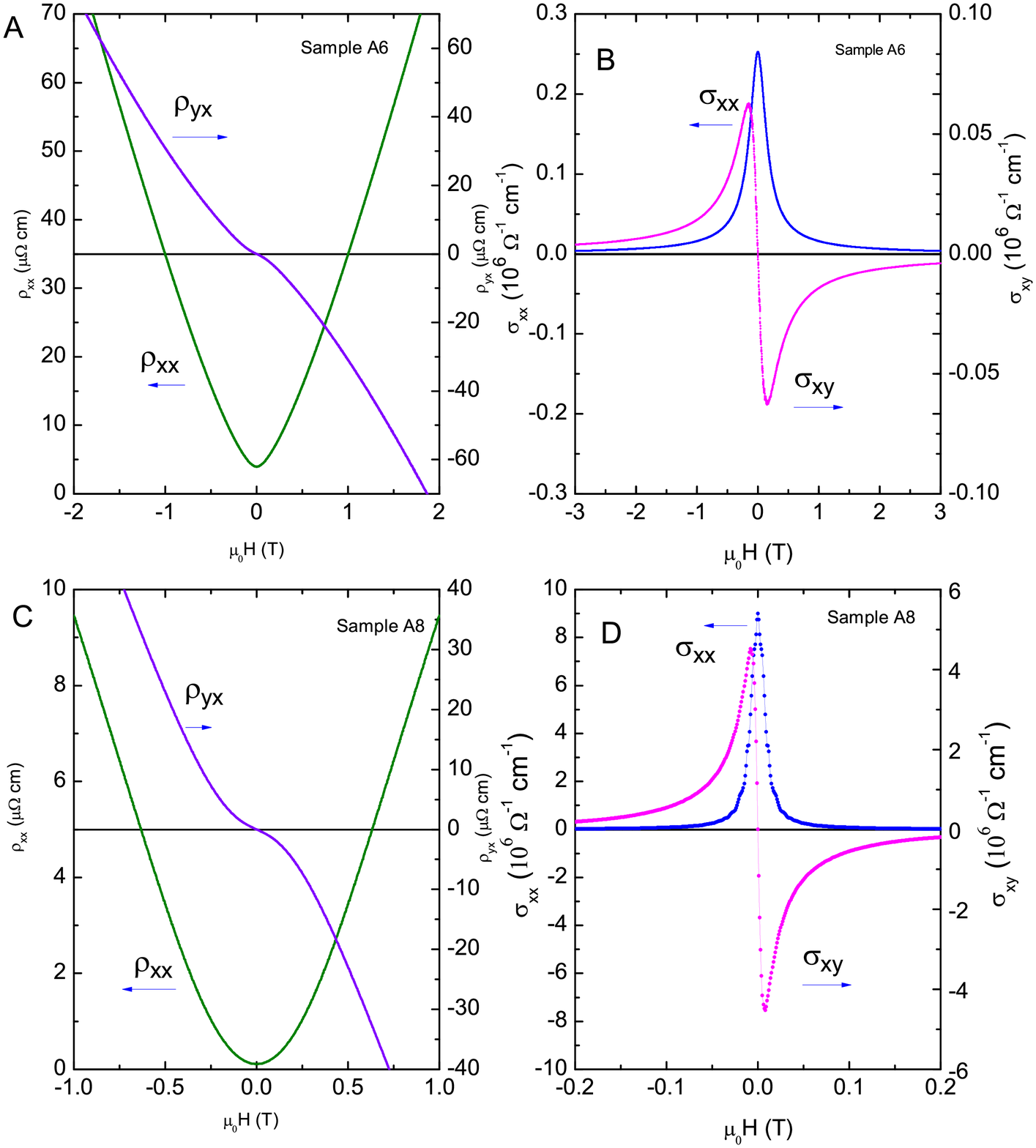}
\caption{\label{figMR68} 
Field profiles of $\rho_{xx}$ and $\rho_{yx}$ measured at 5 K with $\bf H||\hat{z}$ and current $\bf I||\hat{x}$. Panel A shows the resistivity curves in Sample A6. The conductivity curves $\sigma_{xx}(H)$ and Hall conductivity $\sigma_{xy}(H)$ obtained by matrix inversion are plotted in Panel B. The sharp extrema in $\sigma_{xy}$ locate the geometric-mean mobility $\mu_m = \sqrt{\mu_1\mu_2}$. Panel C plots $\rho_{xx}$ and $\rho_{yx}$ at 5 K in Sample A8. The corresponding curves of $\sigma_{ij}(H)$ are in Panel D. Note the factor of 15 difference in the field scale in Panels B and D. 
}
\end{figure}

\begin{figure}[t]
\includegraphics[width=9 cm]{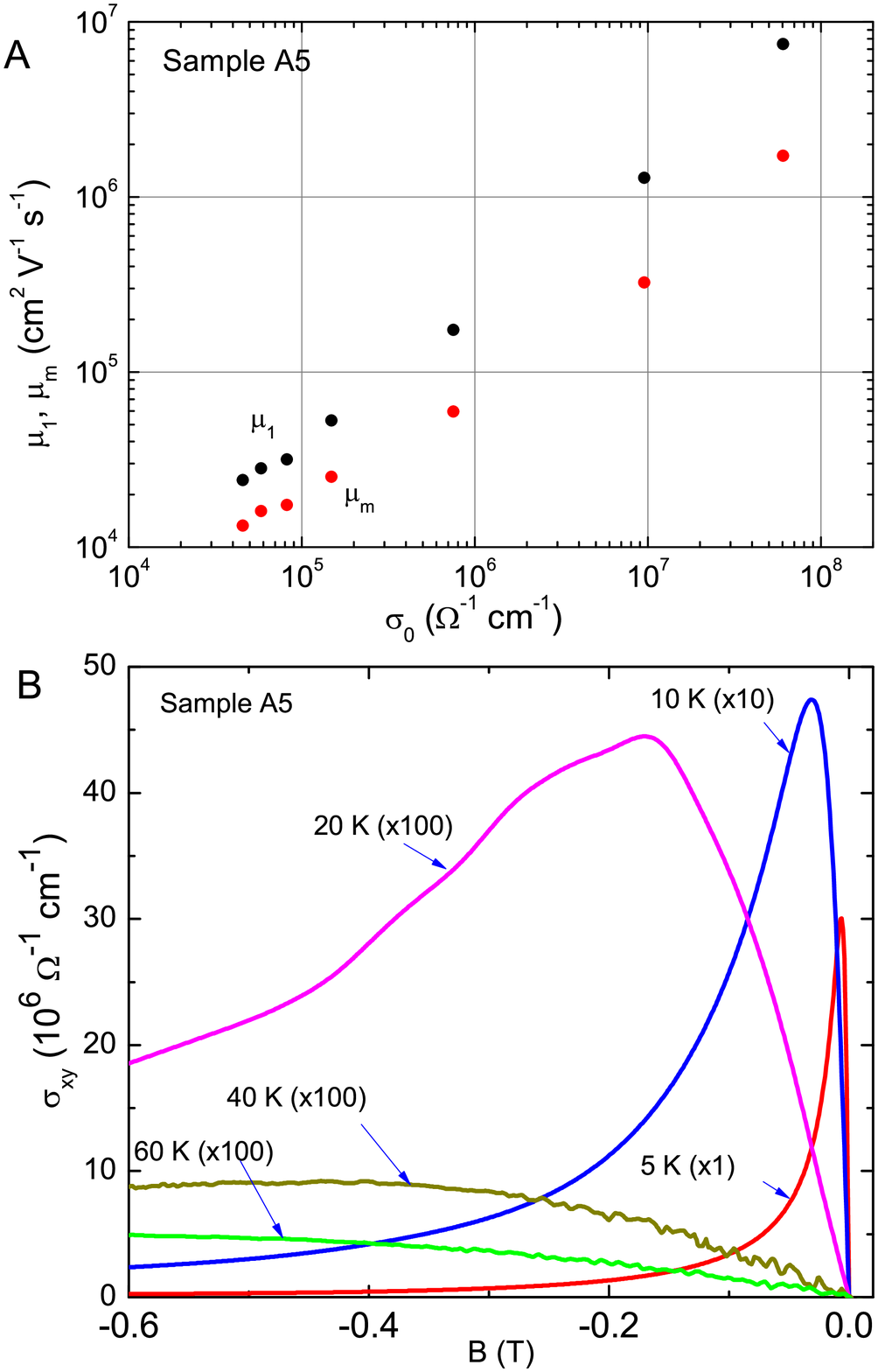}
\caption{\label{figMob} 
Tracking the change in mobility versus $T$ in Sample A5. Panel A plots the mobilities $\mu_1$ and $\mu_m$ versus the zero-field conductivity $\sigma^0_1$ in A5 with $T$ as the parameter. As $T$ is lowered to 5 K, values of $\mu_m$ inferred from the peak field in $\sigma_{xy}$ track very well the increase in $\sigma^0_1$; the mobility $\mu_1$ -- calculated from $\mu_m(T)\sqrt{\gamma(T)}$ -- attains a value close to $10^7$ cm$^2$/Vs. Panel B shows the curves of $\sigma_{xy}(H)$ at selected $T$ from 5 to 60 K. Because of the large variation in peak values of $\sigma_{xy}$, each curve has been multiplied by the vertical scale factor indicated. The peak fields, equal to $\mu_m^{-1}$, shift very rapidly to very small values as $T$ decreases to 5 K.
}
\end{figure}

\section{Dispersive resonant profile of $\sigma_{xy}(H)$}
As discussed in the main text, to determine the geometric mean of the mobilities $\mu_m = \sqrt{\mu_1\mu_2}$, we first measure the curves $\rho_{xx}(H)$ and $\rho_{yx}(H)$. The matrix $\rho_{ij}$ is then inverted to yield the curves $\sigma_{xx}(H)$ and $\sigma_{xy}(H)$. The curves for A4 (lowest mobility) and A5 (highest) were shown in the main text. Here we display in Fig. \ref{figMR68} the curves for the two samples with moderately high mobilities, A6 and A8. Comparing the MR in Panels A and C, we note that the MR in A6 begins to deviate from the $H$-linear profile seen in A4 (and Set B samples), displaying noticeable curvature even at weak $H$. In A8 (with higher mobility still), the curvature is more pronounced. The increased curvature strongly enhances the MR ratio (measured at 9 T) from 35 in A4 to 112 in A6 and 404 in A8 (see Table 1 in main text). In Panels B and D, the curves of $\sigma_{xy}(H)$ display the ``dispersion-resonance'' profile as discussed, with sharp peaks at fields which locate the value $1/\mu_m$. Going from A6 (Panel B) to A8 (Panel D), the mobility $\mu_1$ increases by a factor of 12.5. This causes the peaks to move in by the same factor (note the difference in field scales).

To check that the peak field in $\sigma_{xy}$ accurately measures the mobility $\mu_m$, we can follow the peak field as $T$ is increased in one sample. Figure \ref{figMob}A shows that both $\mu_m(T)$ and $\mu_1(T)$ measured in Sample A5 track closely its zero-$H$ conductivity $\sigma^0_1$ as $T$ varies from 100 K to 5 K ($\mu_1 = \mu_m\sqrt{\gamma}$). Panel B shows the curves of $\sigma_{xy}(H)$ for selected $T$ between 5 and 60 K. The peak values of $\sigma_{xy}$ vary by over 2 orders of magnitude in this range of $T$. Hence, we have multiplied each curve by an appropriate scale factor to make them resolvable.
\vspace{3mm}

\begin{figure}[t]
\includegraphics[width=8 cm]{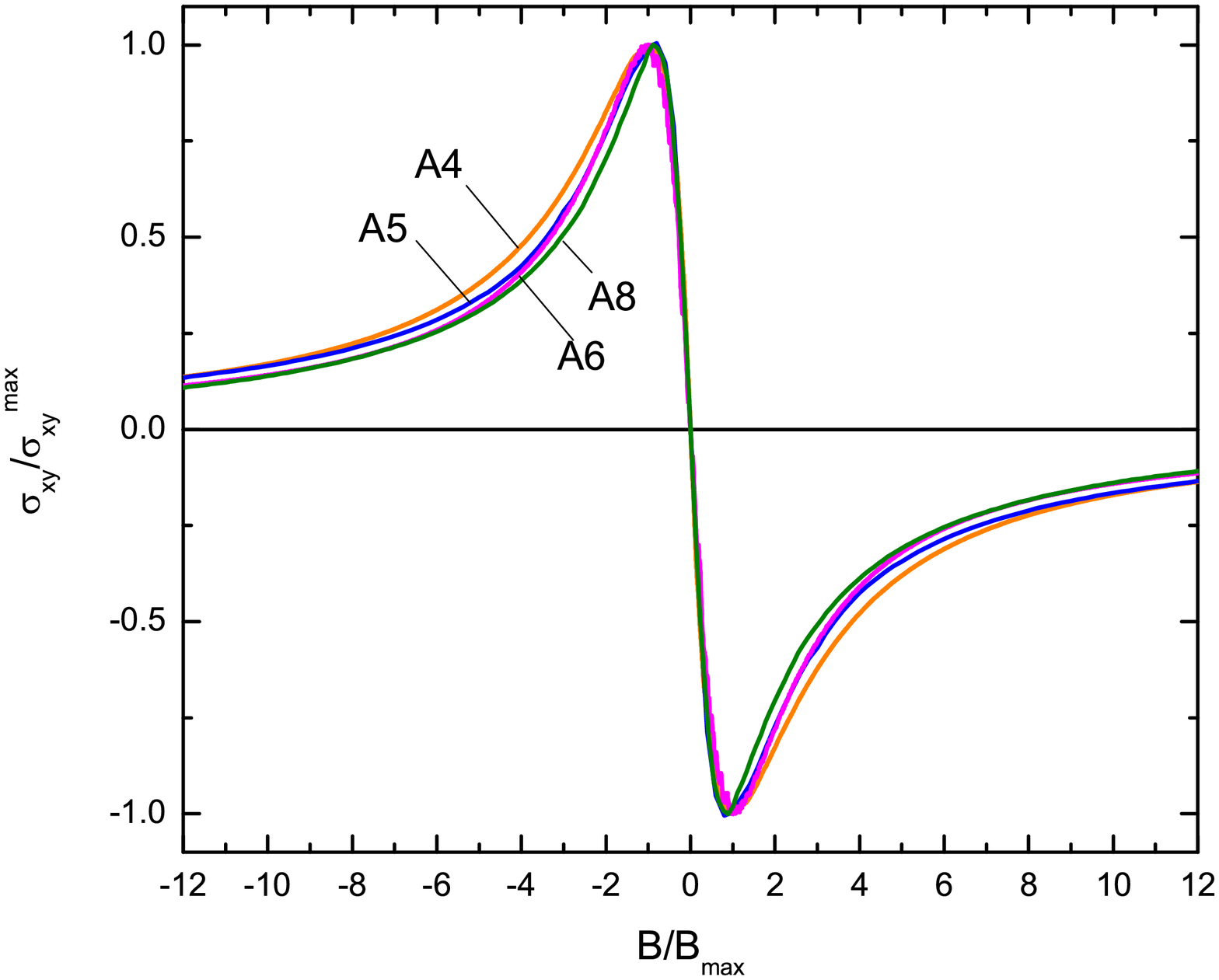}
\caption{\label{figscale} 
Comparison of the normalized Hall conductivity $\sigma_{xy}(H)/\sigma_{xy}^{max}$ vs. the scaled field $B/B_{max}$ in Samples A4, A5, A6 and A8. The scaled form of the Hall conductivity is nominally similar despite a 100-fold change in both $B_{max}$ and $\sigma_{max}$. 
}
\end{figure}

\noindent\emph{Non-uniformity concern and scaling plots}\\
A concern is whether the observed low residual resistivity $\rho_1$ (21 n$\Omega$cm) could arise from a strongly inhomogeneous distribution of lifetimes in the sample. The resonant nature of the peaks in $\sigma_{xy}(H)$ can directly address this issue. In analogy with inhomogeneous broadening in NMR, we expect that a broad distribution of lifetimes (hence mobilities) will also broaden the peak in $\sigma_{xy}$ in proportion. If the increase in conductivity $\sigma^0_1$ (from A4 to A5) is caused by having local regions with a very broad distribution of transport lifetimes, one should see a comparable distribution of peak fields contributing to the measured $\sigma_{xy}(H)$ profile. 

This is not observed. In terms of rescaled variables, we plot the normalized curve $\sigma_{xy}(H)/\sigma_{xy}^{max}$ vs. $B/B_{max}$ where $\sigma_{xy}^{max}$ is the Hall conductivity value at the peak field $B_{max}$ for 4 Set A samples. Within the experimental uncertainty, the widths collapse to the same curve despite a 100-fold change in $B_{max}$. The evidence is that, as $\mu_m$ increases 100-fold (as tracked by the peak), the form of the $\sigma_{xy}$ profile remains unchanged after appropriate rescaling of the field axis; the distribution of mobilities remains very narrow. We argue that this is direct evidence against a broad distribution of lifetimes appearing in the high-mobility samples.

\section{High mobility: comparison with bismuth and 2DEG in GaAs/AlGaAs}
It is interesting to compare the ultrahigh values attained by the mobility $\mu_1$ ($\sim 9\times 10^6$ 
cm$^2$/Vs) with mobilities in in the purest bismuth samples and in the best samples of 2DEG confined in GaAs/AsGaAs quantum wells. There is some spread in the reported mobility values in Bi because both the mobilities of the electron and hole pockets ($\mu_n$ and $\mu_p$) are highly anisotropic. Including values along the 3 axes, there are altogether 6 values of the mobilities to be determined. This is done by fitting extensive magnetoresistance measurements to a Boltzmann-equation model~\cite{Lawson,Bhargava,Hartman}. Most reports obtain values $\mu_n$ in the range 1-10 million cm$^2$/Vs. The highest value is 90 million cm$^2$/Vs reported from a fit by Hartman~\cite{Hartman}. (By contrast, the values in Cd$_3$As$_2$ reported here are directly measured from the peaks in $\sigma_{xy}$ as explained above.) 

The mobilities in 2DEG in GaAs/AlGaAs heterostructures are more reliably determined experimentally since the single FS is isotropic and the carrier density is known to very high accuracy. As reported in Ref. \cite{Pfeiffer}, the highest value is 36 million cm$^2$/Vs. 

A key point is that, in the ``ideal'' Cd$_3$As$_2$ lattice, there are 64 sites for the Cd ions in each unit cell, and that exactly $\frac14$ of the sites are vacant. From the large variation of the RRR (20$\to$4,100) in Set A crystals extracted from the same boule, we infer that the vacancy sites are ordered with a correlation length $\xi$ that varies strongly across crystals grown under nominally similar conditions (the largest RRR is obtained with the longest $\xi$). This disorder leads to strong reduction of the quantum lifetime $\tau_Q$ derived from damping of the SdH oscillations. We remark that neither Bi nor the 2DEG in GaAs/AlGaAs suffer from this type of vacancy disorder (Bi has only 2 atoms per unit cell). Thus it is noteworthy that $\mu_1$ in Cd$_3$As$_2$ attains values nearly comparable to the mobilities in the best Bi and 2DEG samples.

As described in the main text, the transport lifetime $\tau_{tr}$ (which determines the mobility) can be longer than $\tau_Q$ by factors of $R_{\tau}\simeq 10^4$ in Cd$_3$As$_2$. The lattice disorder leads to predominantly forward scattering, but has nearly no effect in relaxing the forward drift velocity. To us, this suggests the existence of strong protection against backscattering, by an unknown mechanism. In the case of 2DEG in GaAs/AlGaAs, $R_{\tau}$ is also very large (10$^2$), but the reason there is now well-understood~\cite{Paalanen,Harrang,Coleridge}. By delta-doping, the doped charge impurities are set back from the 2DEG by 1 micron. The gentle residual disorder seen by electrons in the 2DEG only causes small-angle scattering.

\section{SdH fits and search for a second band}

\begin{figure}[t]
\includegraphics[width=8 cm]{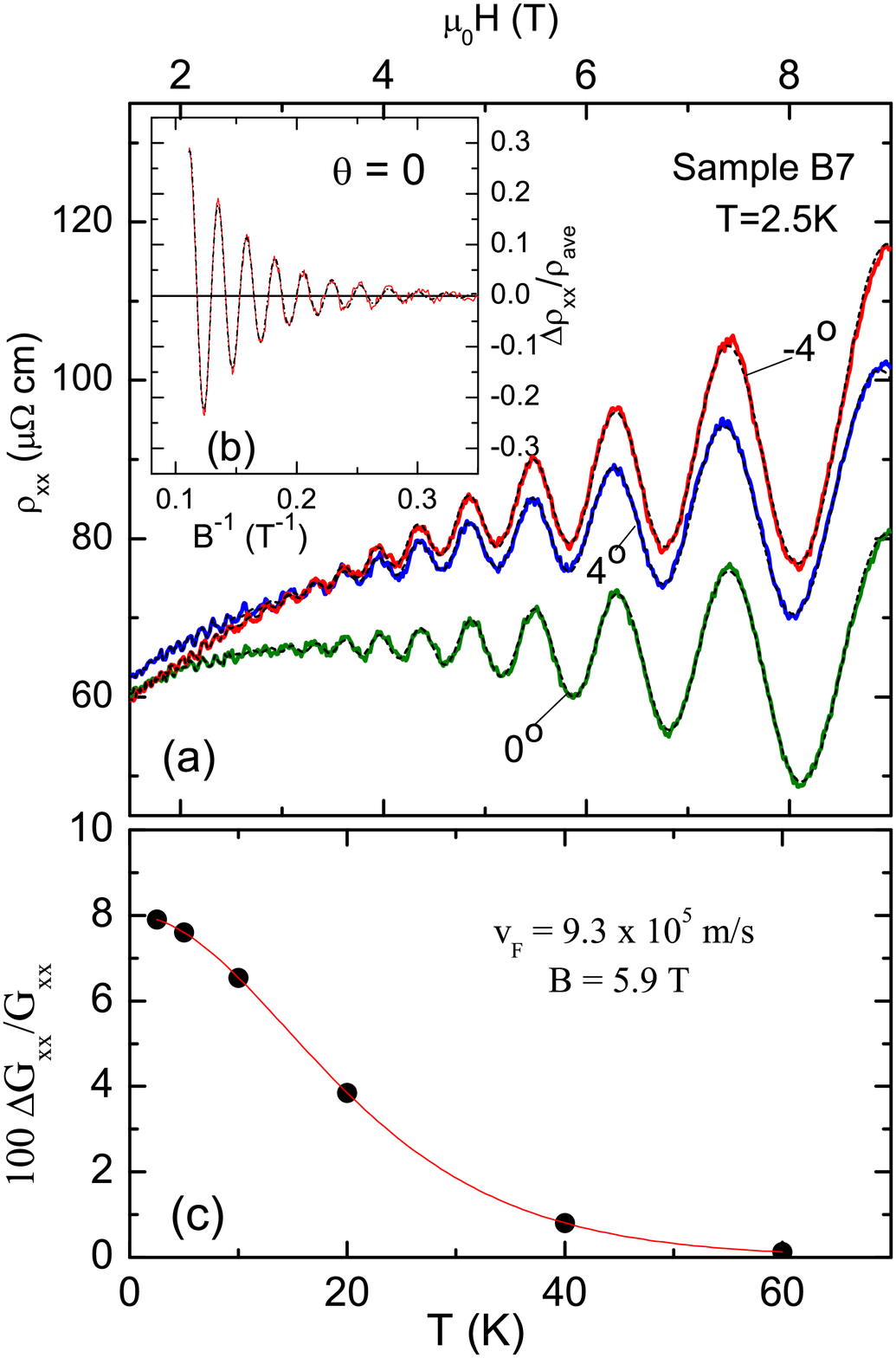}
\caption{\label{figmass}
(Panel a) Fits of $\rho_{xx}$ vs. $H$ to the LK expression (dashed curves) at selected $\theta$ (0, $\pm 4^\circ$) in Sample B7 at 2.5 K. Panel (b) plots the curve and fit at $\theta = 0$ versus $1/B$ to show the exponential damping of the amplitude. (Panel c) The fit to the amplitude at $B$ = 5.9 T vs. $T$ yields the effective mass (or equivalently, the Fermi velocity $v_F$ = 9.3$\times 10^5$ m/s). 
}
\end{figure}

\begin{figure}[t]
\includegraphics[width=8 cm]{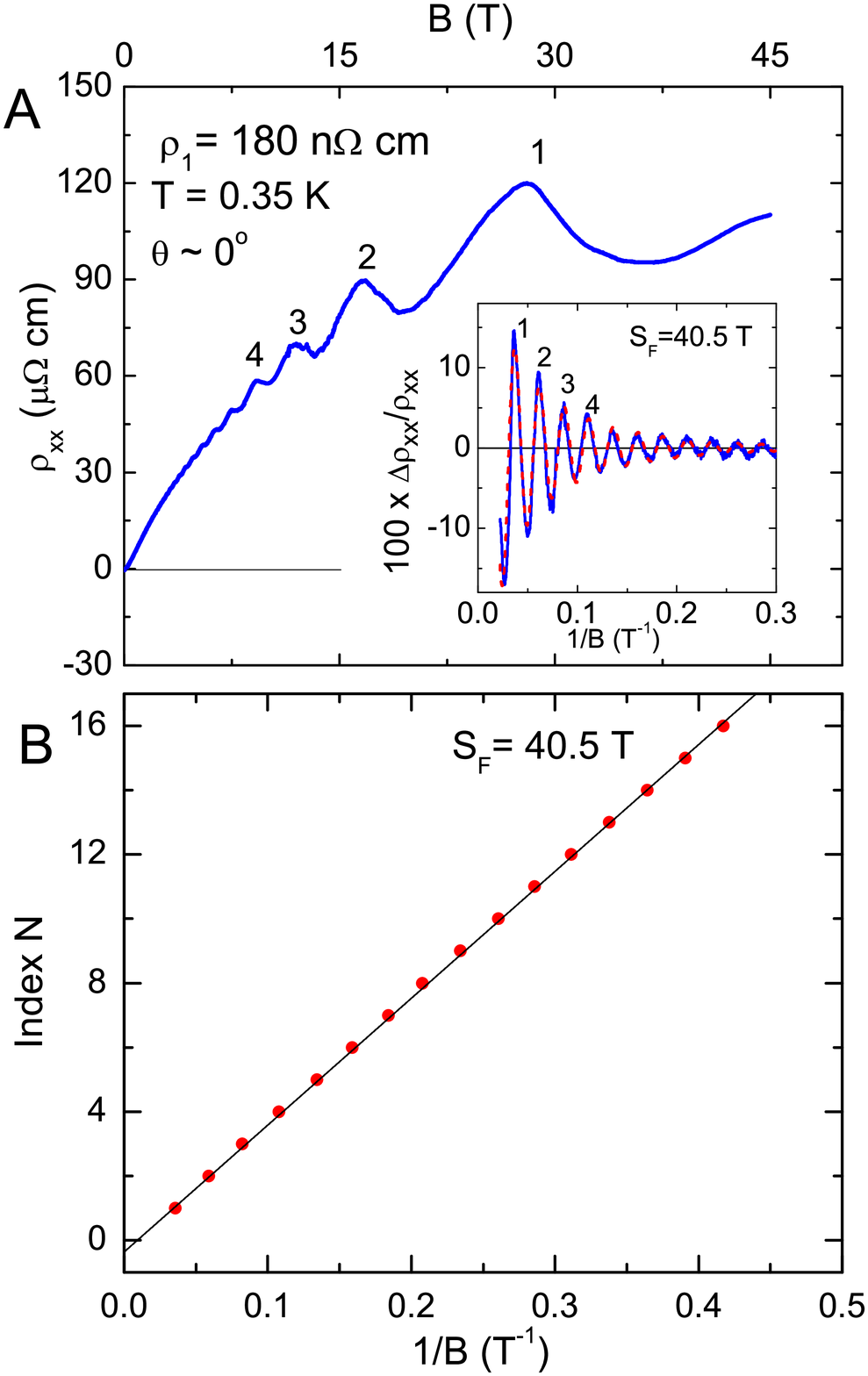}
\caption{\label{fig45T}
High-field SdH oscillations in Cd$_3$As$_2$. Panel A plots the trace of the longitudninal MR $\rho_{xx}$ vs. $H$ up to field 45 T taken in Sample A12 at $T$ = 0.35 K (with $\theta = 0^\circ \pm 5^\circ$). The integers indicate the LL index inferred from Panel B. The inset shows the plot of the ratio $\Delta\rho_{xx}(H)/\rho_{xx}$ (solid curve) after a smooth background curve is subtracted to isolate the SdH oscillations. From the fit to the LK expression (dashed curve) we obtain a quantum lifetime $\tau_Q = 8.56\times 10^{-14}$ s. Panel B is the index plot identifying the index of each Landau level from the maxima in $\Delta\rho_{xx}$. The N=1 level is reached at 27 T. The existence of a second high-mobility band with a different period can be excluded to a resolution of 3$\%$. 
}
\end{figure}

Figure \ref{figmass} plots the SdH curves together with fits to the standard Lifshitz-Kosevich expression in Sample B7 for $\theta=$ 0, 4$^\circ$, -4$^\circ$. We used the Lifshitz-Kosevich (LK) expression in the form
\be
\dfrac{\Delta G_{xx}}{G_{xx}} = \left(\dfrac{\hbar\omega _c}{2E_F}\right)^{1/2} \frac{\lambda}{\sinh\lambda}  e^{-\lambda_D} \cos\left(\frac{\hbar \pi k_F^2}
{eB} +\varphi\right)
\label{LK}
\ee 
with $\lambda = 2\pi^2k_BT/\hbar\omega _c$ and $\lambda _D=2\pi^2k_BT_D/\hbar\omega _c$, where $\omega _c = eB/m_c$ is the cyclotron frequency, with $m_c$ the cyclotron mass. The Dingle temperature is given by $T_D =\hbar /(2\pi k_B\tau _Q)$, with $\tau _Q$ the quantum lifetime. For the Dirac dispersion in Cd$_3$As$_2$, the cyclotron mass is given by $m_c=E_F/v_F^2$. 

To isolate the oscillatory component in each curve of $\rho_{xx}$ vs. $H$, we first determine the strongly $H$-dependent ``background'' $\rho_{ave}$ by averaging out the sinusoidal oscillations (the curve $\rho_{ave}$ at $\theta=0$ is plotted in Fig. 4d in the main text). After this background is removed, we are left with the purely sinusoidal component which is exponentially damped vs. $1/H$ (inset in Fig. \ref{figmass}). This may be fitted to the LK expression by a least-squares fit routine. In general, very close fits to all the observed curves are achieved, as shown (after restoring the background) in Fig. \ref{figmass} for Sample B7 (see Fig. 4B in the main text for A1). The lower panel shows the $T$ dependence of the SdH oscillation amplitude at $B$ = 5.9 T. By fitting to Eq. \ref{LK} (red curve), we determine $m_c$. From the fits, we obtain $S_F$= 44 T, $E_F$ = 220 meV, $v_F$ = 9.3$\times$10$^5$ m/s, $\tau _Q$ = 5.1$\times$10$^{-14}$ s. 
\vspace{3mm}

\noindent\emph{Is there a second band?}\\
To check whether a second band of carriers is present, we have extended the SdH measurements to DC fields of 45 T. Results on a new Set A sample A12 (residual resistivity $\rho_1$ = 180 n$\Omega$cm in zero $H$) are shown in Fig. \ref{fig45T}. Panel A plots the longitudinal MR taken at $T$ = 0.35 K in a field nominally along the needle axis ($\theta = 0^\circ\pm 5^\circ$). SdH oscillations are strongest in the longitudinal MR geometry. After subtracting a smooth background, the SdH oscillations can be fit to the LK expression to yield $S_F$ = 40.5 T (inset). Panel B shows the ``index'' plot of the integers $N$ vs. $1/B_N$ where $B_N$ are the fields at which $\Delta\rho_{xx}$ attains a maximum. The $N$ = 1 level is reached at 27 T. For fields above the last minimum (at 36 T), we begin to access the $N$=0 LL (quantum limit).

If another FS pocket is present with mobility exceeding $\sim 2\times 10^3$ cm$^2$/Vs, its SdH peaks should be visible in the traces of $\rho_{xx}$ and $\Delta\rho_{xx}$, and display a period (vs. $1/B$) distinct from the dominant oscillation, especially in fields above 30 T. To a resolution 3$\%$ of the amplitude of the dominant oscillation, we do not resolve a second band or FS pocket. 

The STM experiment~\cite{Yazdani}, which resolves peaks in the density of states of LLs in fields up to 14 T using QPI (quasiparticle interference), was performed on a crystal extracted from the same flux boule. The QPI results also see only one electron band. Based on results from the two very different experiments on samples extracted from the same growth boule, we can exclude the existence of a second FS pocket distinct from the dominant one.

\begin{figure}[t]
\includegraphics[width=8 cm]{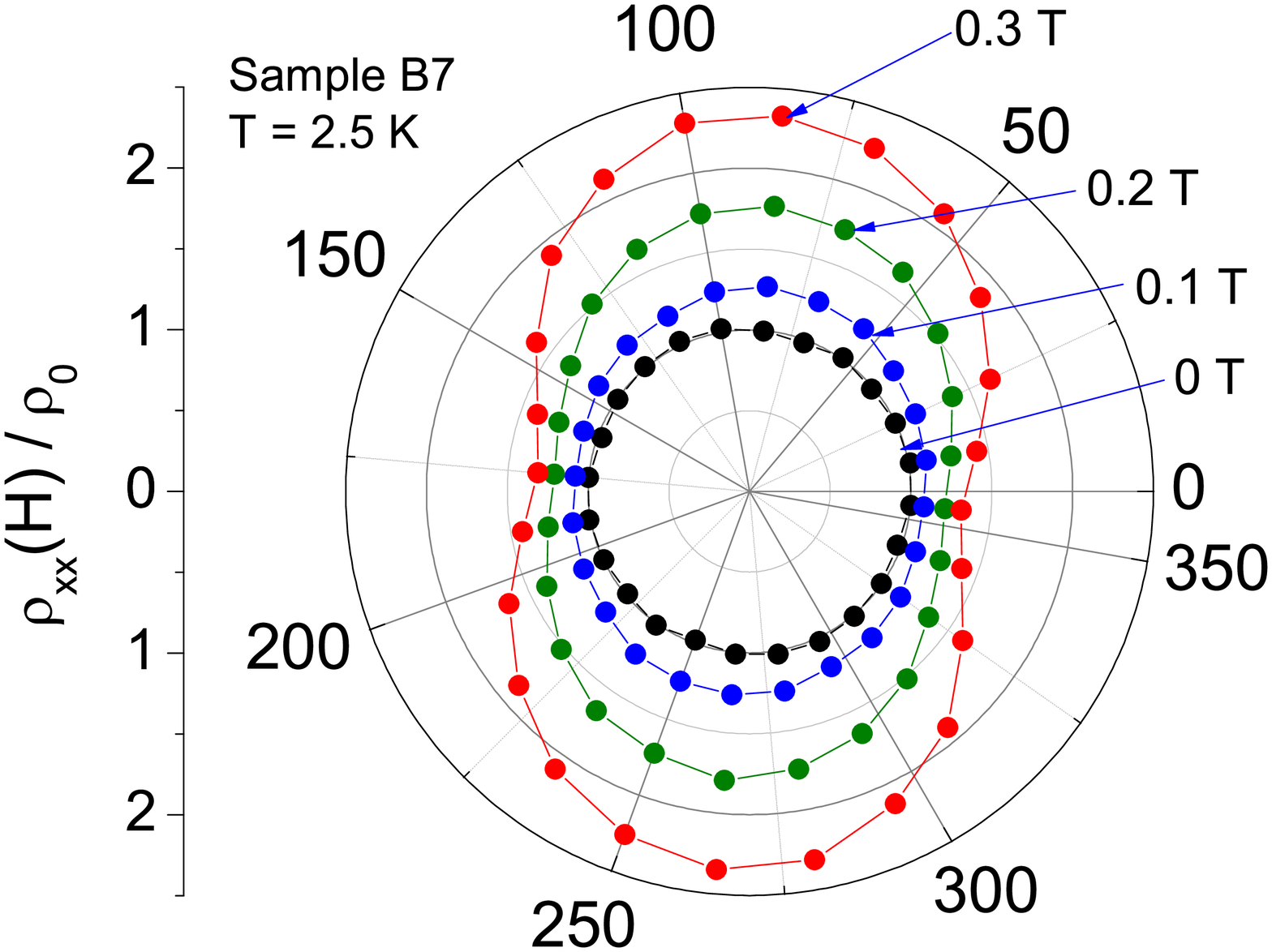}
\caption{\label{figPolarS7}
Polar plot of the angular variation of the low-$T$ MR (in Sample B7) for $H$ fixed at values 0.1,$\cdots$, 0.3 T. The radial coordinate represents the ratio $\rho_{xx}(H)/\rho_0$ (with values shown on the left axis), and the angular coordinate is the tilt angle $\theta$ of $\bf H$. As $H\to 0$, the MR becomes isotropic, suggesting that the weak-$H$ MR is mediated by the spin degrees. 
}
\end{figure}

\begin{figure}[t]
\includegraphics[width=8 cm]{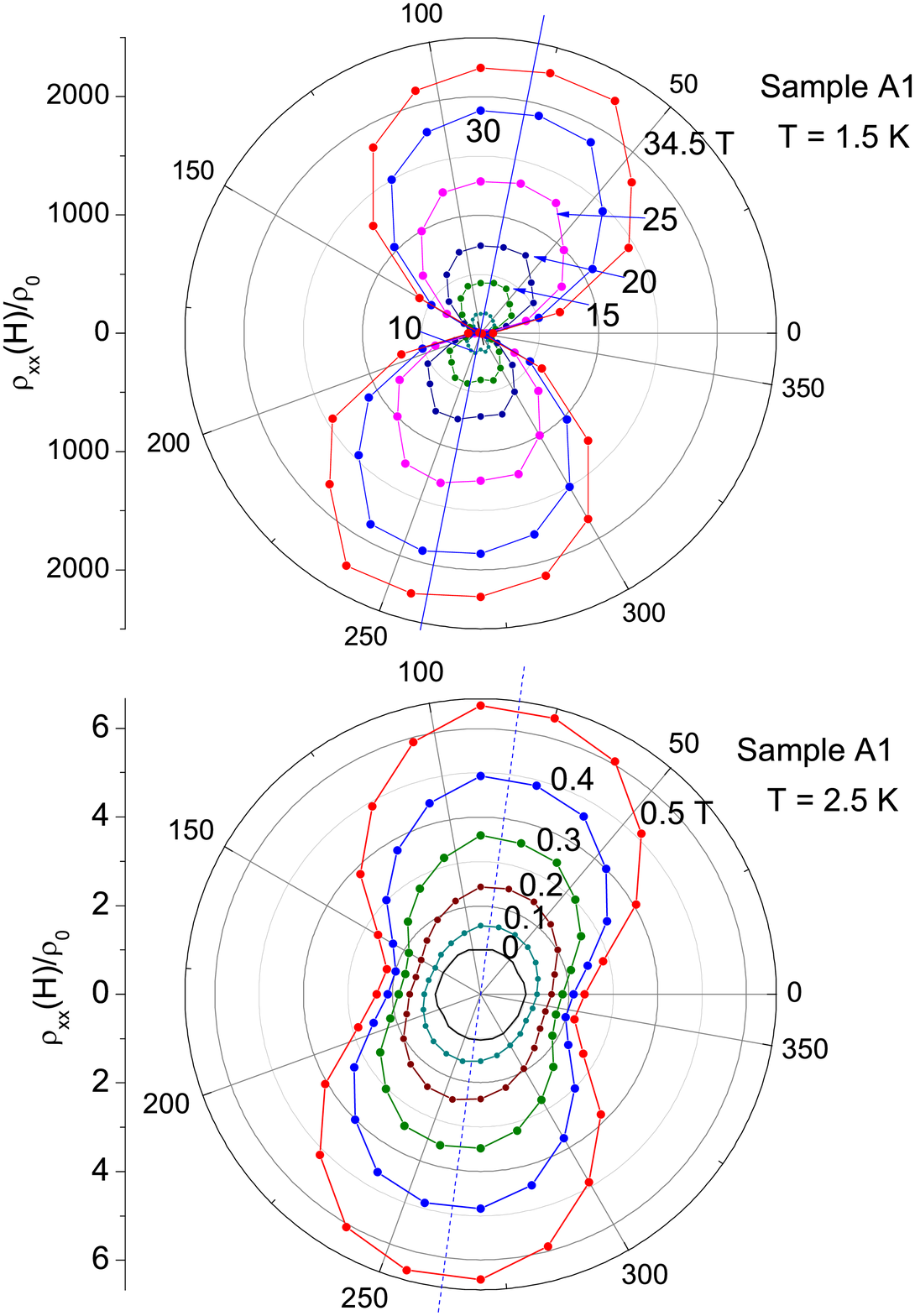}
\caption{\label{figPolar10} 
Polar plot of the angular variation of the low-$T$ MR (Sample A1) in high fields, 10-34.5 T (upper panel) and in very weak $H$ ($\le$0.5 T, lower panel). The radial coordinate represents the ratio $\rho_{xx}(H)/\rho_0$ (with values shown on the left axis). The angular coordinate is the tilt angle $\theta$ of $\bf H$. For $H<$0.1 T, the MR is nearly isotropic, but acquires a dipolar component that grows rapidly as $H$ increases. The weak-field, isotropic regime in Set A samples is confined to very weak $H$, and difficult to investigate compared with Set B samples. The high-field polar plot was acquired 3 months after the low-field results. There is a factor of 3.45 difference between the two polar plots because $\rho_0$ increased from 32 to 110 n$\Omega$cm presumably from aging processes during the intervening period (see text).
}
\end{figure}

\begin{figure}[h]
\includegraphics[width=8 cm]{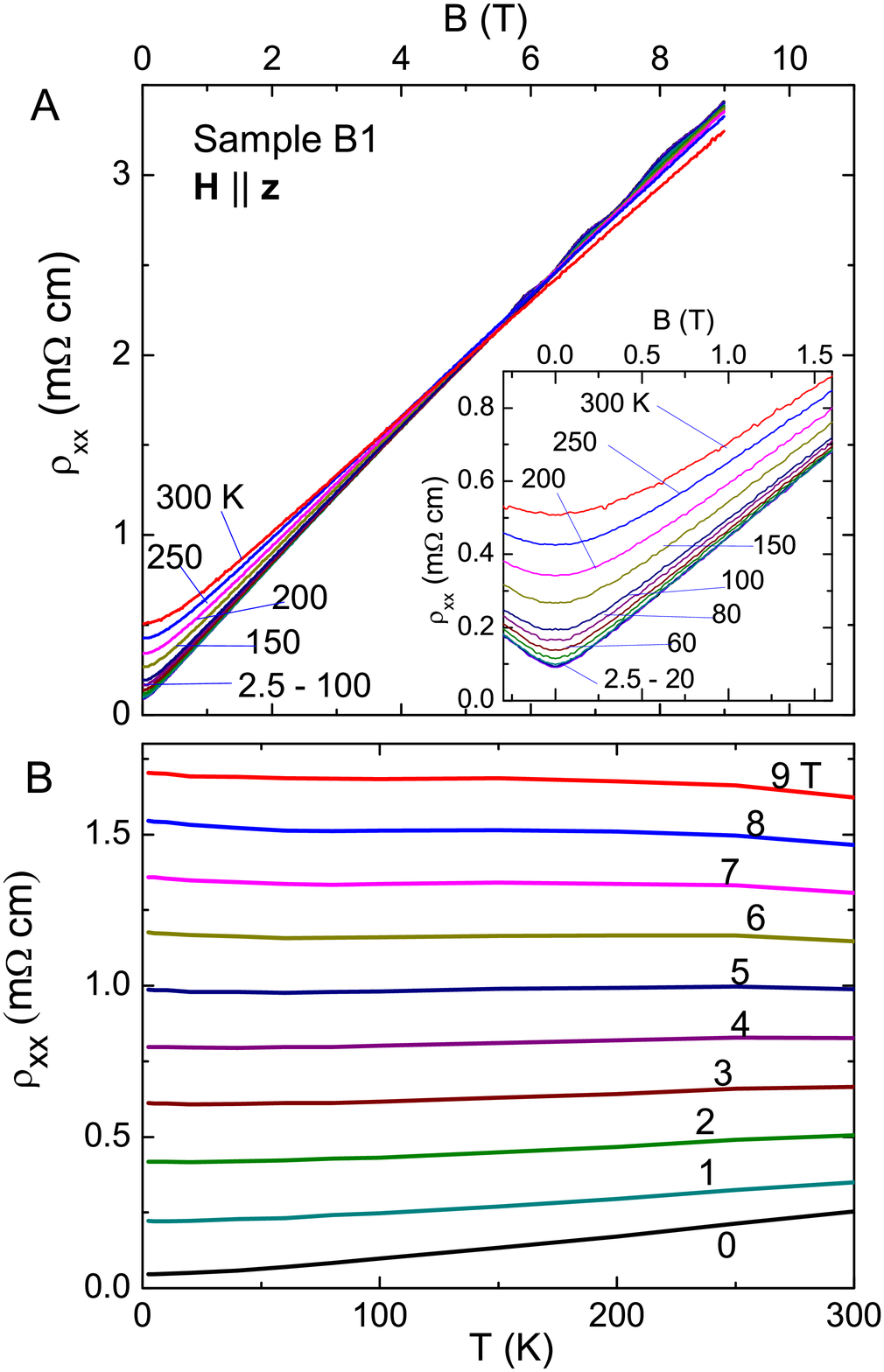}
\caption{\label{figMRvsT} The effect of varying $T$ (2.5$\to$ 300 K) on the MR in Sample B1.
Panel A plots the MR curves $\rho_{xx}$ vs. $H$ at fixed $T$ with $\theta = 90^\circ$. An expanded view of the low $H$ region is given in the inset. 
In Panel B, we plot the $T$ dependence of $\rho_{xx}(T,H)$ with $H$
fixed at selected values. For $H>$ 2 T, the curves are nearly $T$ independent. 
}
\end{figure}

\begin{figure}[h]
\includegraphics[width=9 cm]{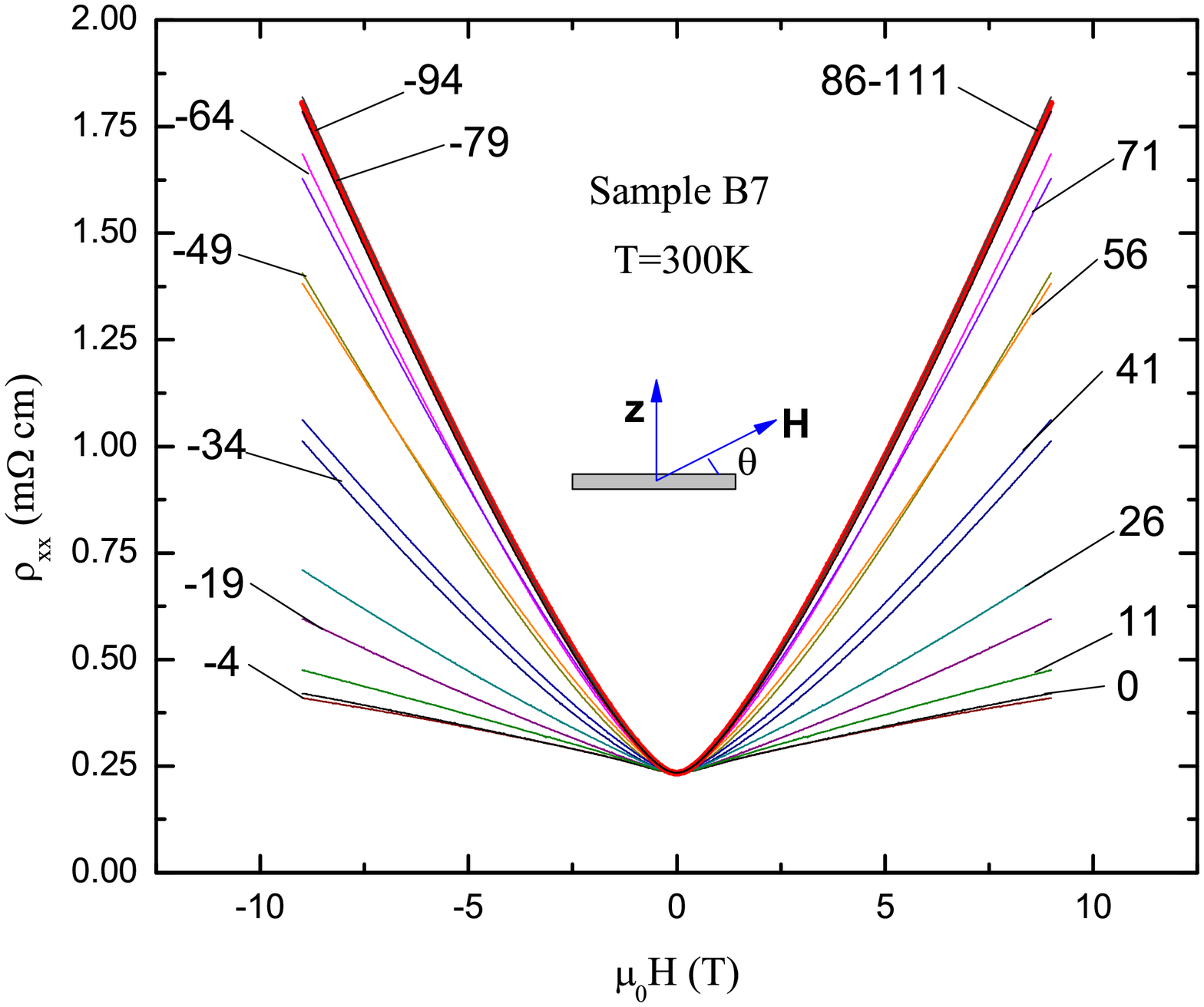}
\includegraphics[width=9 cm]{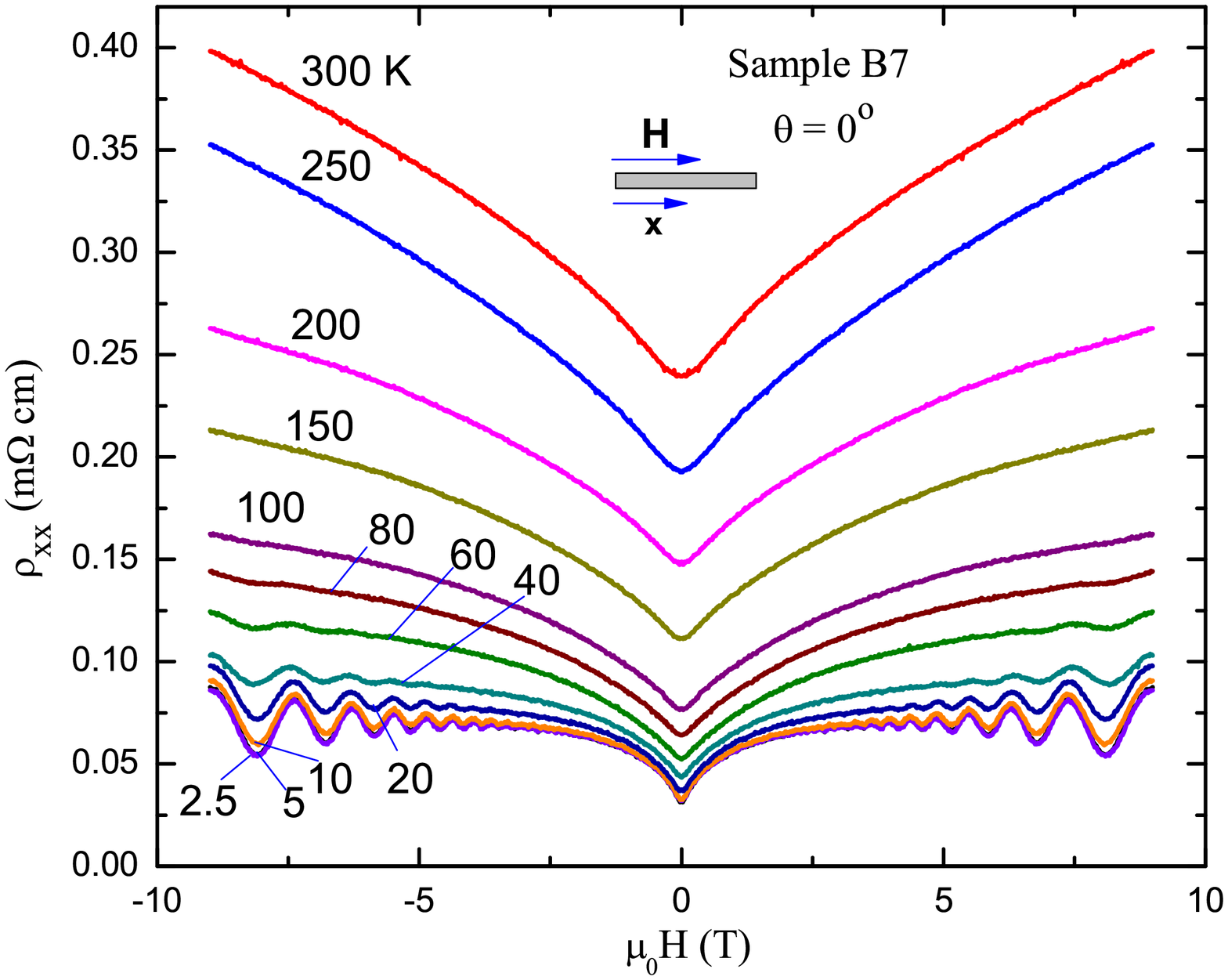}
\caption{\label{figMR300} 
Panel a (upper panel) plots the MR curves $\rho_{xx}$ vs. $H$ at selected field-tilt angles $\theta$ at 300 K
in Sample B7. As $T$ increases from 2.5 to 300 K, the zero-$H$ value of $\rho_{xx}$ increases by a factor of 7.7. However, at large $H$ (e.g. 9 T), $\rho_{xx}\simeq 1.8\, \mathrm{m}\Omega$cm is nearly the same as at 2.5 K. The broadening at $H$=0 is closely similar to that observed in Sample B1. In Panel b (lower panel), we plot the longitudinal ($\theta = 0^\circ$) MR curves as $T$ is reduced from 300 K to 2.5 K. Below 60 K, SdH oscillations become resolvable. As $T\to$2.5 K, a negative MR term can be resolved as a slight decrease in the background curve (when the SdH oscillations are averaged out).
}
\end{figure}

\section{Chiral anomaly}

An interesting prediction in Weyl semimetals is the effect of the ``chiral anomaly'' on transport~\cite{Qi,Nielsen,Son}. In a magnetic field $\bf B||\hat{z}$, each Dirac node is predicted to split into two Weyl nodes with opposite chirality $\chi = \pm 1$. The lowest Landau level (LL) in each Weyl node displays a linear dispersion along $\bf \hat{B}$ with the sign fixed by $\chi$, i.e. the energy of the lowest LL is given by $E_0=-\chi \hbar v_Fk_z$ where $v_F$ is the Fermi velocity. In an electric field $\bf E$, charge $Q$ is pumped between the two branches at the rate $\dot{Q} =-V\frac{e^3}{4\pi^2\hbar^2}\mathbf{E}\cdot \mathbf{B}$, with $V$ the volume of the sample~\cite{Qi}. 

In the quantum limit (only lowest LL occupied), the charge pumping yields a conductivity increment given by~\cite{Nielsen,Son}
\be
\delta\sigma_{zz} = \frac{1}{(2\pi)^2} \frac{e^3}{\hbar^2} Bv_F\tau_v,
\label{szz}
\ee
where $\tau_v$ is the intervalley relaxation time.
Using the mobility $\mu = ev_F\tau_{tr}/\hbar k_F$, we may write the ratio of the chiral term to the zero-$H$ conductivity as
\be
\frac{\delta\sigma_{zz}}{\sigma(0)} = \frac34 \frac{1}{(k_F\ell_B)^2} \frac{\tau_v}{\tau_{tr}},
\ee
where $\ell_B = \sqrt{\hbar/eB}$ is the magnetic length.
For Sample B7, the ratio comes out to $\sim (B/100)(\tau_v/\tau_{tr})$. As a rough estimate of the size of the contribution, we may crudely assume that $\tau_v\sim \tau_{tr}$. The chiral term then gives a negative MR of about 1 $\%$ at 1 T.

\section{Polar plots}

Figure \ref{figPolarS7} shows the polar plots of Sample B7 at low field ($<$ 0.3~T). The observed MR becomes nearly isotropic below 0.3 T. This implies that the MR is mediated by the spin degrees through the Zeeman term. Above 0.3 T, the contribution of the orbital degrees become increasingly important, and the polar plot assumes a dipolar pattern. 

A similar crossover is seen in Sample A1. However, because of the high mobility, the crossover occurs at weaker $H$ ($<$ 0.1 T). Figure \ref{figPolar10}a shows the dipolar pattern in high fields ($H>$ 10 T). In the limit of weak $H$ (Panel b), the pattern crosses over to a nearly isotropic form below 0.1 T.

\section{Temperature dependence and longitudinal MR}
The large transverse MR extends to 300 K. Figure \ref{figMRvsT}A plots curves of $\rho_{xx}(T,H)$ in Sample B1 ($\theta = 90^\circ$) at several temperatures. As $T$ is increased from 2.5 to 300 K, the MR profile is nominally unchanged except that the parabolic variation in weak $H$ becomes more evident (we show the minima on expanded scale in the inset). We observe that $\rho_{xx}(T,H)$ is nominally $T$ independent above 2 T. The effect of $T$ is pronounced in weak $H$ but becomes insignificant at large $H$. In Panel B, we have replotted the data in Panel A as $\rho_{xx}$ vs. $T$ with $H$ fixed at selected values. Whereas at $H = 0$, the profile is strongly metallic, it rapidly becomes $T$ independent when $H$ exceeds 2 T. This behavior should be contrasted with what is observed in Bi where the fixed-field curves retain strong $T$ dependence even when $H$ is very large.

In Fig. \ref{figMR300}a, we keep $T$ fixed at 300 K (data from Sample B7), but rotate $\theta$ from 90$^\circ$ to $0^\circ$. As discussed in the main text, the MR ratio measured at 2.5 K is strongly suppressed when $\theta\to 0$. The pattern at 300 K is similar (except that the minimum at $H$ = 0 is significantly rounded as the mobility decreases). 

Finally, we show how raising $T$ affects the small negative MR contribution observed in a longitudinal $\bf H$ (see $\rho_{ave}$ in Fig. 4B of the main text). With $\theta$ fixed at 0, we warm up the sample to 300 K. The SdH oscillation amplitude is suppressed above 40 K. Significantly, above 20 K, the negative MR term rapidly becomes unresolvable. At 300 K, the longitudinal MR is strongly positive. These plots show that the negative MR is a low-temperature feature that is easily suppressed above 20 K.

{We thank Andrei Bernevig, Sid Parameswaran, Ashvin Vishwanath and Ali Yazdani for valuable discussions, and Nan Yao for assistance with the EDX. N.P.O. is supported by 
the Army Research Office (ARO W911NF-11-1-0379). R.J.C. and N.P.O. are supported by funds from a MURI grant on Topological Insulators (ARO W911NF-12-1-0461) and the US National Science Foundation
(grant number DMR 0819860). T.L acknowledges scholarship support from the Japan Student Services Organization. Some of the experiments were performed at the National High Magnetic Field Laboratory, which is supported by National Science Foundation Cooperative Agreement No. DMR-1157490, the State of Florida, and the U.S. Department of Energy.}

\end{document}